\documentclass[aps,reprint,nofootinbib,nobibnotes,notitlepage,superscriptaddress,onecolumn,prd,
 amsmath,amssymb
]{revtex4-1}

\usepackage[caption=false]{subfig}
\usepackage{braket}
\usepackage{float}
\usepackage{lipsum}
\usepackage{graphicx}
\usepackage{dcolumn}
\usepackage{bm}
\usepackage{natbib}
\usepackage{hyperref}
\usepackage[capitalise]{cleveref}
\hypersetup{
	colorlinks = true,
    linkcolor = Red,
    urlcolor  = Red,
    citecolor = Red
}

\usepackage{microtype}

\usepackage{verbatim}
\usepackage[amssymb]{SIunits}
\usepackage{tabularx}
\usepackage[dvipsnames]{xcolor}
\usepackage{wasysym}

\usepackage{tikz}

\def\be{\begin{equation}}
\def\ee{\end{equation}}

\usepackage{color}
\definecolor{darkgreen}{RGB}{0,120,0}
\definecolor{darkblue}{RGB}{20,120,120}

\newcommand{\Mpch}{h^{-1}\mathrm{Mpc}}
\newcommand{\hMpc}{h\,\mathrm{Mpc}^{-1}}

\newcommand{\delD}[1]{(2\pi)^3\delta_\mathrm{D}\left({#1}\right)}

\newcommand{\av}[1]{\left\langle{#1}\right\rangle} 

\newcommand{\vk}{\vec k}
\newcommand{\hk}{\hat{\vec k}}
\newcommand{\vK}{\vec K}
\newcommand{\vp}{\vec p}

\newcommand{\vq}{\vec q}
\newcommand{\vx}{\vec x}
\newcommand{\hx}{\hat{\vec x}}
\newcommand{\hy}{\hat{\vec y}}
\newcommand{\vy}{\vec y}

\newcommand{\hn}{\hat{\vec n}}

\newcommand{\tjo}[3]{\begin{pmatrix} {#1} & {#2} & {#3}\\ 0 & 0 & 0\end{pmatrix}}
\newcommand{\tj}[6]{\begin{pmatrix} {#1} & {#2} & {#3}\\ {#4} & {#5} & {#6}\end{pmatrix}}

\renewcommand{\vr}{\vec r}

\def\beq{\begin{eqnarray}}
\def\eeq{\end{eqnarray}}

\let\vec\mathbf

\makeatletter
\newlength{\apb@width}
\newcommand{\autoparbox}[2][c]{\settowidth{\apb@width}{#2}\parbox[#1]{\apb@width}{#2}}

\makeatother

\def\mnras{\rm{MNRAS}}

\def\physrep{\rm{Phys.~Rep.}}
\def\apj{\rm{ApJ}}                 
\def\apjs{\rm{ApJS}}               
\def\jcap{\rm{JCAP}}

\begin{document}

\title{Signatures of a Parity-Violating Universe}

\author{William R. Coulton}
\email{wcoulton@flatironinstitute.org}
\affiliation{Center for Computational Astrophysics, Flatiron Institute, 162 5th Avenue, New York, NY 10010, USA}

\author{Oliver H.\,E. Philcox}
\email{ohep2@cantab.ac.uk}
\affiliation{Center for Theoretical Physics, Columbia University, New York, NY 10027, USA}
\affiliation{Simons Society of Fellows, Simons Foundation, 160 5th Avenue, New York, NY 10010, USA}

\author{Francisco Villaescusa-Navarro}
\email{fvillaescusa@flatironinstitute.org}
\affiliation{Center for Computational Astrophysics, Flatiron Institute, 162 5th Avenue, New York, NY 10010, USA}
\affiliation{Department of Astrophysical Sciences, Princeton University, 4 Ivy Lane, Princeton, NJ 08544 USA}

\begin{abstract} 
    What would a parity-violating universe look like? We present a numerical and theoretical study of mirror asymmetries in the late universe, using a new suite of $N$-body simulations: \textsc{quijote-odd}. These feature parity-violating initial conditions, injected via a simple ansatz for the imaginary primordial trispectrum and evolved into the non-linear regime. We find that the realization-averaged power spectrum, bispectrum, halo mass function, and matter PDF are not affected by our modifications to the initial conditions, deep into the non-linear regime, which we argue arises from rotational and translational invariance. In contrast, the parity-odd trispectrum of matter (measured using a new estimator), shows distinct signatures proportional to the parity-violating parameter, $p_{\rm NL}$, which sets the amplitude of the primordial trispectrum. We additionally find intriguing signatures in the angular momentum of halos, with the primordial trispectrum inducing a non-zero correlation between angular momentum and smoothed velocity field, proportional to $p_{\rm NL}$. Our simulation suite has been made \href{https://quijote-simulations.readthedocs.io/en/latest/odd.html}{public} to facilitate future analyses.
\end{abstract}

\maketitle

\section{Introduction}

\noindent Parity describes the transformation properties of a system under point reflections; roughly speaking, a symmetric universe is one that looks identical when left and right are exchanged. From a terrestrial viewpoint, violations of this symmetry are commonplace: examples include neutrinos (which are solely left-handed), amino acid chirality, and the handedness of the brain. Physically, many of these effects are sourced by the weak interaction, which, as discovered empirically in 1957 \citep{Wu:1957my,PhysRev.104.254}, does not obey parity-symmetry (unlike other standard model forces), but rather the more general symmetry of charge-parity-time conservation. 

Zooming out, one may ask whether parity-symmetry should be broken also on cosmological scales. In this instance, physics is dominated not by the weak force but by gravitation, which is invariant under reflections (in the Einsteinian paradigm). As such, a bound on large-scale parity-asymmetries provides a probe of the Universe's initial conditions. In the simplest models of inflation, single-field slow-roll, parity symmetries are obeyed \citep{Cabass:2022rhr}; however, violation can occur as a result of exotic physics, which would indicate phenomena such as new forces or interactions with novel particles \citep{Bordin:2020eui,1999PhRvL..83.1506L,2011JCAP...06..003S,2015JCAP...01..027B,2015JCAP...07..039B,2017JCAP...07..034B,2017JCAP...07..034B,2016PhRvD..94h3503S,Biagetti:2020lpx,2010PhRvD..81l3529G}. Indirect evidence for mirror asymmetries may already exist; the known asymmetry of baryons over antibaryons requires a charge-parity- and charge-violating process, as described by the Sakharov conditions \citep{1967JETPL...5...24S}, which could occur via some form of gravitational parity-violation \citep{2009PhR...480....1A,2013JCAP...04..046A,2016IJMPD..2540013A,2006PhRvL..96h1301A}.

Ascertaining whether the Universe is parity-symmetric is a topic of particular current relevance. Using data from large spectroscopic surveys, recent works have revealed a slight asymmetry in the distribution of chiral tetrahedra of galaxies \citep{Philcox:2022hkh,Hou:2022wfj}, using a method first presented in \citep{2021arXiv211012004C}. Whilst it is important to bear in mind that various systematic effects could cause this result, there remains an intriguing possibility that this signal is physical. As discussed above, this could indicate parity-violating processes at work in inflation or at late-times (e.g., via some flavor of chiral gravity), though the latter requires an exceedingly large characteristic length scale \citep{Cabass:2022oap}. As shown in \citep{Philcox:2023ffy}, the former explanation is also disfavored, since the relevant signal does not show up in the cosmic microwave background trispectrum, where it would be expected to arise at some $50\sigma$. This suggests that systematics (mischaracterization of noise in particular), may be to blame. Future data will shed much more light on such results, and there remains the possibility that robust signatures could be detected with future surveys such as DESI, Euclid, MegaMapper and beyond \citep[e.g.,][]{2016arXiv161100036D,2011arXiv1110.3193L,Schlegel:2022vrv}.

If an early-Universe source of parity-violation exists, how best can it be searched for? Put another way: how do mirror asymmetries in inflation manifest themselves in the distribution of matter and galaxies today? Previous observational studies have principally focused on the four-point correlation function or trispectrum (though see also \citep{Motloch:2020qvx,Yu:2019bsd}), which, in the linear regime, directly probes the (parity-sensitive) primordial trispectrum of inflation. At late-times however, the Universe is significantly more complex due to non-linear gravitational evolution. Does this leak information into lower-point functions? Or change the masses of galaxies? Or their angular momenta? To probe such effects, we require numerical codes to simulate gravitational effects down to redshift zero and thus assess the impact on cosmological statistics such as power spectra, halo mass functions, and beyond.

\begin{figure}
    \centering
    \includegraphics[width=\textwidth]{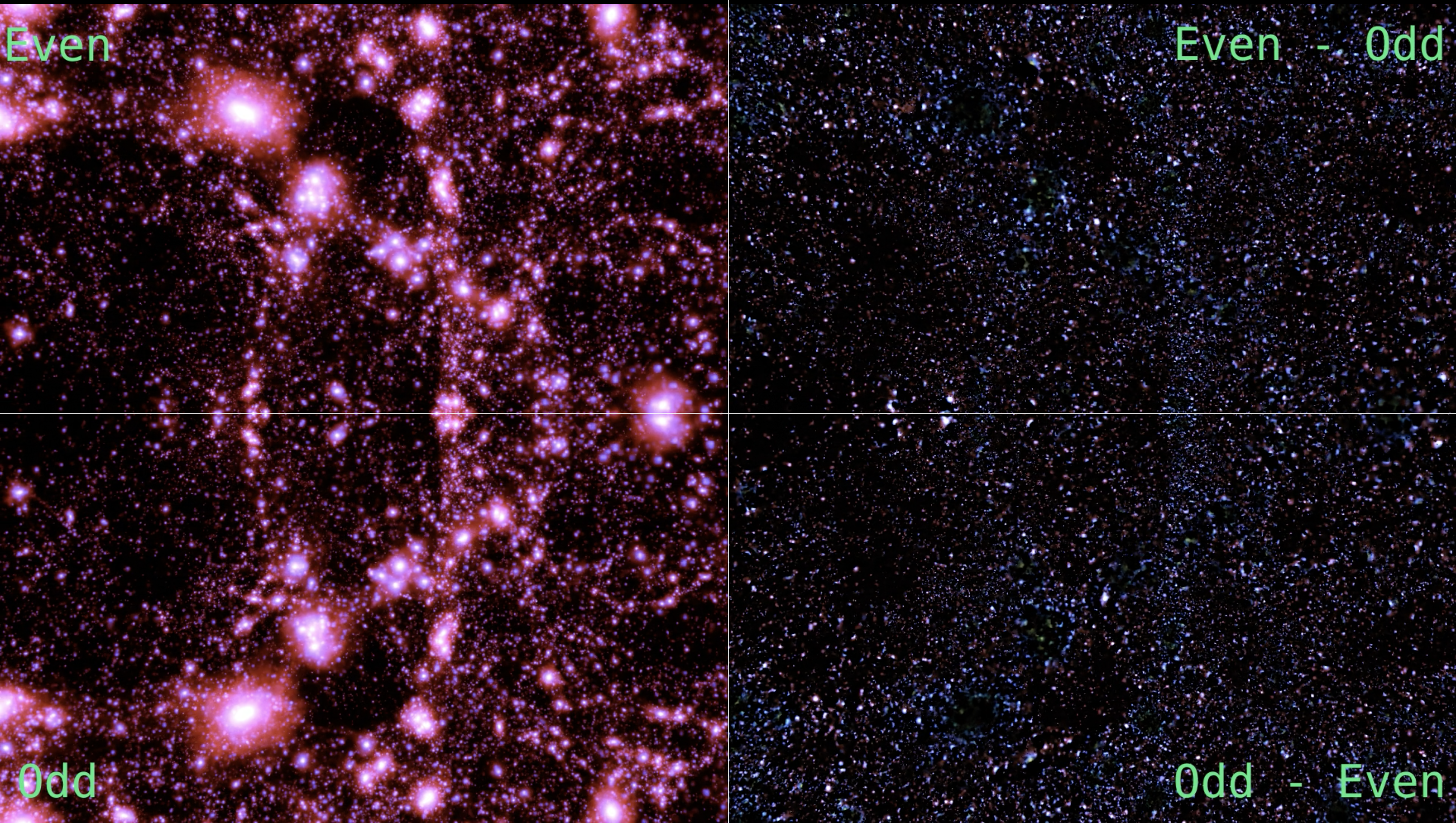}
    \caption{We show two N-body simulations: one with standard initial conditions (top-left) and one with parity-violating initial conditions (bottom-left) with $p_{\rm NL}=10^7$. Both simulations follow the evolution of $512^3$ particles on a periodic volume of $(50~h^{-1}{\rm Mpc})^3$ down to $z=0$, and they share the same underlying Gaussian density field. The parity-violating simulation has been flipped along the horizontal axis to give the impression that there is a mirror at the center of the image. The panels on the right part show the difference between the densities of the two simulations. As can be seen, parity-violating induces non-negligible differences in the distribution of matter in the Universe, the statistical implications of which will be discussed in this work. The images are generated from a full ray-tracing calculation at redshift zero from blender.}
    \label{fig: slice}
\end{figure}

In this work, we will present the first numerical study of cosmologies with primordial scalar parity-violation. To this end, we will consider the generation of asymmetric initial conditions from a fiducial template and generate their low redshift counterparts, creating an extension to the \textsc{quijote} suite \citep{Villaescusa-Navarro:2019bje}, dubbed \textsc{quijote-odd}. By way of introduction, we show a slice through an analogous (zoomed-in) simulation in Fig.\,\ref{fig: slice}, comparing to its parity-even counterpart; an associated movie can be found online.\footnote{Movie available at \href{https://www.youtube.com/watch?v=4bnKGFYoLpA&t=2s&ab_channel=FranciscoVillaescusa-Navarro}{www.youtube.com/watch?v=4bnKGFYoLpA\&t=2s\&ab\_channel=FranciscoVillaescusa-Navarro}} Using these realizations, we can robustly compute the late-time manifestations of parity-violating initial conditions and compare them to theoretical predictions, made possible in part by novel trispectrum estimators. Here, we will consider a range of statistics: the power spectrum, bispectrum, trispectrum, halo mass function, matter PDF, and halo angular momentum statistics. Each will be measured from both parity-conserving and parity-violating simulations, which will allow us to test theoretical expectations and uncover various future avenues for exploration.

\vskip 4pt
The remainder of this work is as follows. We begin with a general introduction to scalar parity-violation, presenting theoretical arguments for the parity-sensitive of various observables in \S\ref{sec: theory}. In \S\ref{sec: ics}, we consider how to generate parity-violating initial conditions, before \S\ref{sec: sims} discusses the associated simulation suite. \S\ref{sec: results} shows results of various correlators extracted from our simulations, and compares to theoretical predictions, before we conclude in \S\ref{sec: conc}, with Appendices \ref{app: estimator}\,\&\,\ref{app: tl-theory} discussing mathematical minutiae pertaining to trispectrum estimation and theory modeling.

\section{Parity-Violation in Theory}\label{sec: theory}

\subsection{Definition}
\noindent Consider a scalar function, $\zeta$, depending on $n$ displacement vectors $\vr_1,\cdots,\vr_n$. The action of a three-dimensional parity-transformation, denoted $\mathbb{P}$, is the following:
\beq
    \mathbb{P}[\zeta(\vr_1,\cdots,\vr_n)] = \zeta(-\vr_1,\cdots,-\vr_n),
\eeq
which is equivalent to a point reflection. If $\mathbb{P}[\zeta] = \pm\zeta$, we describe the statistic as parity-even or parity-odd respectively. Assuming isotropy, the scalar function $\zeta$ can depend only on the position vectors $r^a_i$, derivatives $\partial^a$, and rotationally-invariant quantities, \textit{i.e.}\ $\delta_{\rm K}^{ab}$ and $\epsilon^{abc}$ (for Cartesian indices $a,b,\cdots$). Under the parity-transformation, all vectors pick up a factor of $-1$; this implies that parity-odd correlators \textit{must} contain an odd number of Levi-Cevita symbols, $\epsilon^{abc}$, since there is no other way to combine an odd number of vectors to form a scalar.

In Fourier-space, $\zeta$ depends on some set of wavevectors, $\vk_1,\cdots,\vk_n$ (each associated to a single $\vr_i$), which transform as follows under parity:
\beq
    \mathbb{P}[\zeta(\vk_1,\cdots,\vk_n)] = \zeta(-\vk_1,\cdots,-\vk_n).
\eeq
Often, $\zeta$ is the correlation of some number of real-fields; in this case, $\zeta(-\vk_1,\cdots,-\vk_n) = \zeta^*(\vk_1,\cdots,\vk_n)$, thus parity-even (parity-odd) correlators are purely real (imaginary). Analogously to before, any parity-odd correlator in Fourier-space requires a Levi-Cevita symbol, leading to a term of the form $\vk_i\cdot\vk_j\times\vk_k$ (since $k$-derivatives are killed by locality). 

\subsection{Correlators: Heuristic Argument}
\noindent Physical symmetries place strong restrictions on the parity-sensitivity of various observables. Here, we will consider explicitly the general form of the power spectrum, bispectrum, and trispectrum of real fields, and argue that, in expectation, only the latter is parity-sensitive. Our arguments are non-perturbative, and thus apply in arbitrarily non-linear regimes, though require all observables to be averaged over realizations.

By translation invariance, the two-, three-, and four-point correlators of real fields $X,Y,Z,W$, must satisfy momentum conservation:
\beq
	\av{X(\vk_1)Y(\vk_2)} &=& P_{XY}(\vk_1)\delD{\vk_1+\vk_2}\\\nonumber
	\av{X(\vk_1)Y(\vk_2)Z(\vk_3)} &=& B_{XYZ}(\vk_1,\vk_2,\vk_3)\delD{\vk_1+\vk_2+\vk_3}\\\nonumber
	\av{X(\vk_1)Y(\vk_2)Z(\vk_3)W(\vk_4)} &=& T_{XYZW}(\vk_1,\vk_2,\vk_3,\vk_4)\delD{\vk_1+\vk_2+\vk_3+\vk_4}.
\eeq
Asserting rotational invariance (ignoring, e.g., wide-angle effects), each correlator can depend only on scalar functions derived from the $\vk_i$ wavenumbers, as discussed above. In three-dimensions, these comprise only $|\vk|\equiv k$, $\mu_{ij}\equiv \hk_i\cdot\hk_j$, $\hk_i\cdot\hk_j\times\hk_k$ leading to the definitions:
\beq\label{eq: trans-rot-inv-statistics}
	P_{XY}(\vk_1) &=& P_{XY}(k_1)\\\nonumber
	B_{XYZ}(\vk_1,\vk_2,\vk_3) &=& B_{XYZ}(k_1,k_2,k_3,\mu_{12},\mu_{13},\mu_{23},\hk_1\cdot\hk_2\times\hk_3) = B_{XYZ}(k_1,k_2,k_3)\\\nonumber
	T_{XYZW}(\vk_1,\vk_2,\vk_3,\vk_4) &=& T_{XYZW}(k_1,k_2,k_3,k_4,|\vk_1+\vk_2|,|\vk_1+\vk_3|,\mu_{12},\mu_{13},\mu_{14},\mu_{23},\mu_{24},\mu_{34},\hk_1\cdot\hk_2\times\hk_3)\\\nonumber
	 &=& T_{XYZW}(k_1,k_2,k_3,k_4,|\vk_1+\vk_2|,|\vk_1+\vk_3|,\hk_1\cdot\hk_2\times\hk_3),
\eeq
where we eliminate degenerate variables to reach the second definitions, using conservation of momentum. As above, parity-inversion leads to the mapping $\vk\to-\vk$, and, if the fields are real, a complex conjugation; it is immediately apparent that $P_{XY}$ and $B_{XYZ}$ are invariant under this transformation, and are thus parity-insensitive and real. In contrast, the trispectrum (and any higher-order correlators) contains the irreducible triple-product $\hk_1\cdot\hk_2\times\hk_3$, which allows it to be parity-sensitive, and, in general, complex. 

\subsection{Correlators: Formal Proof}
\noindent Whilst the above argument suffices to demonstrate that (realization-averaged) power spectra and bispectra cannot probe parity-violating physics, due to their lack of dependence on the scalar triple product, some readers may wish for a more formal argument. To this end, let us assume some primordial source of parity-violation, through the gravitational potential $\phi$. By the above arguments, this must involve a scalar triple product, taking the heuristic form
\beq\label{eq: general-phi-Fourier}
    \phi^{\rm NG}(\vk) = \phi^{\rm G}(\vk) + ip_{\rm NL}\int_{\vp_1+\vp_2+\vp_3=\vec k}(\vp_1\cdot\vp_2\times\vp_3)f[\phi^{\rm G}](\vp_1,\vp_2,\vp_3) + \cdots,
\eeq
where $\phi^{\rm G}$ is the Gaussian potential, $\int_{\vp}\equiv \int d^3p/(2\pi)^3$, and the scalar functional $f$ will be discussed in \S\ref{sec: ics}. Here, the magnitude of parity-violation is set by the imaginary parameter $ip_{\rm NL}$. Searching for late-time manifestations of parity-violation is thus equivalent to finding realization-averaged observables that depend on $ip_{\rm NL}$. Notably, those depending on \textit{even} powers of $ip_{\rm NL}$ are not signatures of parity-violation, since the square of a parity-odd contribution is parity-even. For this reason, we will consider simulations with both positive and negative $p_{\rm NL}$ to validate our arguments in the below, since their difference removes any quadratic (and higher) $ip_{\rm NL}$ contributions.

Any late-time field $X(\vk)$ is uniquely determined by the primordial potential $\phi^{\rm NG}$, and can thus be written as a formal Taylor series:
\beq\label{eq: general-expansions}
    X(\vk) = \sum_{n=0}^\infty \int_{\sum_i\vq_i=\vk}X^{(n)}(\vq_1,\cdots,\vq_n;\vk)\phi^{\rm NG}(\vq_1)\cdots  \phi^{\rm NG}(\vq_n),
\eeq
where the $n$-th order piece involves $n$ primordial potentials. The deterministic kernel $X^{(n)}$ is fixed by the equations of motion (independently of inflationary physics); ignoring late-time parity-violation, it can depend only on magnitudes $|\vq_i|$ and angles $\hat{\vq}_i\cdot\hat{\vq}_j$.\footnote{If there were late-time parity-violation, the $X^{(n)}$ kernels would include a parity-odd triple product of the form $(i\sigma)\vq_1\cdot\vq_2\times\vq_3$, for amplitude parameter $i\sigma$. The coupling of this to the $ip_{\rm NL}$ triple product in the initial conditions would source \textit{parity-even} physics linear in $ip_{\rm NL}$ (via $i^2p_{\rm NL}\sigma$). However, since the underlying equations of gravitation and hydrodynamics are parity-even, this cannot be sourced by standard physics at any order.} Expanding the potentials to first order in $ip_{\rm NL}$ leads to the formal contribution to the power spectrum (which remains non-perturbative and could include arbitrary non-linearity, loops, counterterms, and beyond):
\beq
    P_{XY}(\vk) \supset \sum_{n=1}^\infty\int_{\sum_i\vq_i=\vk}\left[ip_{\rm NL}(\vq_1\cdot\vq_2\times\vq_3)\right]\times K^{(n)}_{XY}(q_i,k,\hk\cdot\hat{\vq},\hat{\vq}_i\cdot\hat{\vq}_j)\times [P_\phi(q_i)P_\phi(|\vk-\vq_i|)\cdots],
\eeq
where $K^{(n)}_{XY}$ is the $n$-th order kernel which is built from $X^{(n)}$ and $Y^{(n')}$ and $P_\phi$ is the power spectrum of $\phi^{\rm NG}$. Here, the parity-violating initial conditions source the square bracket, including both $ip_{\rm NL}$ and a scalar triple-product. Under the transformation $\vk\to-\vk$ and a relabelling $\vq_i\to-\vq_i$, we find $P^{(ip_{\rm NL})}_{XY}(-\vk) = -P^{(ip_{\rm NL})}_{XY}(\vk)$, since $K^{(n)}_{XY}$ depends only on magnitudes and angles so is invariant, yet $\vq_1\cdot\vq_2\times\vq_3$ changes sign. Since rotational invariance restricts us to power spectra of the form $P_{XY}(\vk) = P_{XY}(k)$, this implies that the order $ip_{\rm NL}$ contribution must vanish. An analogous argument implies that higher odd powers of $ip_{\rm NL}$ vanish, thus the overall power spectrum is parity-insensitive. By a similar line of reasoning, the bispectrum must also be parity-even; however, the trispectrum can exhibit signatures of parity-violation, since rotational symmetry does not demand it to be independent of $\hk$.

Before continuing, we briefly remark on violations of the above assumptions. The most pertinent of these is the breaking of rotational invariance through redshift-space distortions. In this case, the problem contains an additional vector: the (assumed global) line-of-sight $\hn$. For the power spectrum, symmetry then dictates $P_{XY}(\vk_1) = P_{XY}(k_1,\hk_1\cdot\hn)$. Note that as exchanging $X\leftrightarrow Y$ is equivalent to $\vk_1\to-\vk_1$, a signature of parity-violation can occur only for cross-spectra with $X\neq Y$ (as $P_{XX}(\vk)-P_{XX}(-\vk) = 0$ by definition). Furthermore, parity-violation is equivalent to replacing the line-of-sight $\hn$ with $-\hn$: however, this is an isometry of the equations of motion, thus there again can be no signal proportional to $ip_{\rm NL}$. To see this, note the formal expansion for the redshift-space power spectrum:
\beq
    P_{XY}(k,\hk\cdot\hn) &\supset& \sum_{n=1}^\infty \int_{\sum_i\vq_i=\vk}\left[ip_{\rm NL}(\vq_1\cdot\vq_2\times\vq_3)\right]\times K^{(n)}_{XY}(q_i,k,\hk\cdot\hat{\vq},\hat{\vq}_i\cdot\hat{\vq}_j,\hat{\vq_i}\cdot\hn,\hk\cdot\hn)\\\nonumber
    &&\,\times\, [P_\phi(q_i)P_\phi(|\vk-\vq_i|)\cdots].
\eeq
Here the line-of-sight $\hn$ enters only in the kernels (since it impacts the transformation between observables and initial conditions). Importantly, these kernels are invariant under $\hn\to-\hn$ due to cylindrical symmetry (\textit{i.e.}\ the fact that the line-of-sight is defined only up to $180\degree$ rotations).\footnote{This can be violated with wide-angle effects, since there are then two distinct lines-of-sight. Such an effect cannot be probed with standard $N$-body simulations however.} Since $\hn$ does not couple to the initial conditions, we find that, under $\vk\to-\vk$, $\hn\to-\hn$ and relabelling $\vq\to-\vq$, $P_{XY}^{(ip_{\rm NL})}(k,\hk\cdot\hn)$ transforms to $-P_{XY}^{(ip_{\rm NL})}(k,\hk\cdot\hn)$, thus the contribution must vanish. The same logic applies to the bispectrum, whence parity-violation is equivalent to $\hn\to-\hn$ as before, but the kernels remain invariant under $\hn\to-\hn$.

\subsection{Halo Mass Function}

\noindent Next, we consider the halo mass function, $n(M)$. Once again, this is not parity-sensitive. Formally, an explicit expression for $n(M)$ is given by as a Taylor series in Fourier-space (cf.\,\ref{eq: general-expansions}):
\beq
    \widehat{n}(M) = \sum_{n=0}^\infty \int_{\sum_i\vq_i=\vec 0}N^{(n)}(\vq_1,\cdots,\vq_n,M)\phi^{\rm NG}(\vq_1)\cdots\phi^{\rm NG}(\vq_n),
\eeq
for some kernel $N^{(n)}$. Averaging over the initial conditions, we find the following schematic contribution at first order in $ip_{\rm NL}$
\beq
    n(M) \supset \sum_{n=1}^\infty \int_{\sum_i\vq_i=\vec 0}\left[ip_{\rm NL}(\vq_1\cdot\vq_2\times\vq_3)\right]\times\bar{N}^{(n)}(q_i,\hat{\vq}_i\cdot\hat{\vq}_j,M)\times\left[P_\phi(q_1)\cdots\right],
\eeq
with new kernel $\bar{N}^{(n)}$. Under the relabelling $\vq\to-\vq$, we find $n^{(ip_{\rm NL})}(M)=-n^{(ip_{\rm NL})}(M)$, thus the contribution is zero. The same holds at higher (odd) orders.

\subsection{Halo Angular Momenta}\label{subsec: theory-spin}
\noindent In the above discussion, we have considered only scalar observables, such as the matter density or mass function. Different behavior can arise if one instead works with tensor observables, since these carry additional directional information, bypassing the above symmetry constraints. For the CMB, the principal example is polarization, which allows for parity-sensitive two-point correlators ($TB$ and $EB$ spectra) \citep{1999PhRvL..83.1506L,2010PhRvD..81l3529G,2011JCAP...06..003S,2012JCAP...06..015S,2015JCAP...07..039B,2016PhRvD..94h3503S,2017PhRvL.118v1301M,2017JCAP...07..034B,Planck:2016soo,Orlando:2022rih,2020PhRvL.125v1301M}. In the context of the late Universe, the simplest tensor fields at our disposal are the shapes of galaxies, as well as the velocity and angular momentum fields of tracer particles, such as halos or galaxies. The former are difficult to probe using $N$-body simulations, thus we will restrict our attention to the latter in this work.

Under a parity-transform, velocity transforms as a vector (with $\mathbb{P}[\vec v(\vr)]=-\vec v(-\vr)$), whilst the angular momentum transforms as an axial vector (with $\mathbb{P}[\vec J(\vr)]=\vec J(-\vr)$). As such, the combination $\vec J\cdot\vec v$ is a pseudo-scalar, which changes sign under point reflections. This enables a convenient test: averaging over a sufficiently large number of objects, the cosine $\hat{\vec J}\cdot\hat{\vec v}$ can take non-zero values only if parity symmetry is broken. In the above language, $\hat{\vec J}\cdot\hat{\vec v}$ must be proportional to odd powers of the parity-breaking amplitude $ip_{\rm NL}$. We will use this notion to probe parity violation in \S\ref{sec: results}.

Furthermore, one can decompose (axial) vectors such as the angular momentum into a left- and right-handed components, and thus evaluate their chirality. This was demonstrated in \citep{Yu:2019bsd} (and later works \citep{Motloch:2021lsw,Motloch:2021mfz,Motloch:2020qvx,Jia:2022sys}), considering the quantity $\mu_{L,R}\equiv \hat{\vec J}\cdot\hat{\vec J}_{L,R}$, where $\hat{\vec J}$ is the observed halo angular momentum, and $\hat{\vec J}_{L,R}$ is a proxy constructed from the primordial density field. From the sum $\mu_L+\mu_R$, one can probe if galaxy angular momentum is correlated with the initial conditions; from the difference, one can probe parity violating processes. This will again be explored in \S\ref{sec: results}.

\subsection{Summary}
\noindent The conclusion of the above discussion is that the parity violation is difficult to observe when considering only scalar observables, such as the mass and halo density fields. At arbitrarily non-linear scales, the power spectrum, bispectrum, and halo mass function are unaffected by parity-violation injected in the initial conditions, \textit{i.e.}\ they are independent of the sign of $p_{\rm NL}$. As such, suites of simulations with $+|p_{\rm NL}|$ and $-|p_{\rm NL}|$ should yield the same averaged statistics in arbitrarily non-linear regimes; a prediction that we verify in \S\ref{sec: results}. To probe parity-violation with scalar observables, we require quantities which depend on \textit{at least three} independent vector coordinates (e.g.,\,$\vk_1,\vk_2,\vk_3$), the simplest of which is the parity-odd trispectrum. 

In contrast, tensorial observables can probe parity violation with simpler statistics. A particularly notable example is the angular momentum of tracers, which probes the mirror asymmetries both via its helicity decomposition, and its scalar product with the velocity field.\footnote{It is interesting to consider whether the angular momentum field can impart parity-sensitivity on power spectra (and beyond) via observational selection effects, such as $\delta \propto \hn\cdot\hat{\vec J}$ (\textit{i.e.}\ a tendency to observe more face-on than edge-on galaxies). Since the statistics of the angular momentum field are themselves isotropic and there is only one global axis in the problem ($\hn$), any contractions with the Levi-Cevita symbol must vanish, thus such phenomena cannot affect matter correlators.} Such quantities may have limitations in practice however, since we have access only to projected velocity fields, and can measure only the tangential angular momentum, through galaxy spin \citep[e.g.][]{Motloch:2021lsw}. Though beyond the scope of this work, $EB$ galaxy shear correlations may also be an intriguing probe of parity violation.

\section{Parity-Violation in Practice: Initial Conditions}\label{sec: ics}

\subsection{Definition}
\noindent We now consider how to create simulations with injected parity-violation. Analogously to the case of three-point non-Gaussianities (proportional to $f_{\rm NL}$) \citep{Scoccimarro:2011pz}, our procedure is to first generate a Gaussian primordial potential, $\phi^{(1)}(\vx)$, then modulate it to source the correlators of interest. Definitively, we perform the following transformation:
\beq\label{eq: parity-transform}
  {\phi}^{\rm NG}(\vx)= \phi^{(1)}(\vx) \to \phi^{(1)}(\vx) + p_{\rm NL}\left[\epsilon^{ijk}(\partial_i|\partial|^\alpha\phi^{(1)})(\partial_j|\partial|^\beta\phi^{(1)})(\partial_k|\partial|^\gamma\phi^{(1)})\right](\vx),
\eeq 
where $\alpha\neq\beta\neq\gamma$ are integers, $p_{\rm NL}A_s$ controls the amplitude of parity-violation (for inflationary amplitude $A_s$), and $|\partial|^\alpha$ corresponds to a Fourier-space multiplication by $k^\alpha$. As discussed in \S\ref{sec: theory}, this is parity-violating due to the presence of a Levi-Cevita symbol, which must be contracted with three different fields. If the initial conditions are to be scale-invariant, we require a trispectrum scaling as $(P_\phi)^3$: this is achieved by setting $\alpha+\beta+\gamma=-3$.\footnote{This can also be implemented by acting on the above with $|\partial|^{-\alpha-\beta-\gamma-3}$.} In the below, we will set $\{\alpha,\beta,\gamma\} = \{-2,-1,0\}$. Notably, \eqref{eq: parity-transform} does not represent all possible parity-violating initial conditions, much as the local bispectrum shape does not represent all types of three-point primordial non-Gaussianity. Different models of new physics generate distinct primordial signatures which can be more complex than the above; the only generic prediction is that the modification to the potential must contain an odd number of Levi-Cevita symbols (to ensure parity-violation). 

Defining the Fourier-space field by $\phi(\vk) \equiv \int d\vx\,e^{-i\vk\cdot\vx}\phi(\vx)$, the correction term in \eqref{eq: parity-transform} can be written:
\beq\label{eq: parity-fourier}
    \phi^{(3)}(\vk) = ip_{\rm NL}\int_{\vp_1\vp_2\vp_3}\delD{\vk-\vp_1-\vp_2-\vp_3}\left[\vp_1\cdot\vp_2\times\vp_3\right]p_1^\alpha p_2^\beta p_3^\gamma \phi^{(1)}(\vp_1)\phi^{(1)}(\vp_2)\phi^{(1)}(\vp_3),
\eeq
as in \eqref{eq: general-phi-Fourier}. Practically, this is easiest to compute as a summation in real-space, having first computed the gradient fields $\partial_i|\partial|^\alpha\phi^{(1)}$ via Fourier transforms. We stress that such simplifications are possible only due to our assumption of a separable form for the primordial parity-odd distortion.

The above modification to the primordial potential generically leads to corrections to the power spectrum at $\mathcal{O}(p_{\rm NL}^2)$, which become relevant on small scales.\footnote{As discussed in \S\ref{sec: theory}, rotation and translation invariance forbids any $\mathcal{O}(p_{\rm NL})$ corrections from appearing. Since the residual corrections do not depend on the sign of $p_{\rm NL}$, they are not parity-sensitive.} These may be optionally removed by rescaling the non-Gaussian initial conditions via
\beq\label{eq: IC-rescaling}
\phi^\mathrm{NG}(\vk) \to \sqrt{\frac{\av{\phi^{(1)}(\vk) {\phi^{(1)}}^*(\vk)}}{\av{\phi^\mathrm{NG}(\vk) { \phi^\mathrm{NG}}^*(\vk)}}} \phi^\mathrm{NG}(\vk),
\eeq
which ensures that the output power spectrum is invariant, upon averaging over realizations. Since this factor is $1+\mathcal{O}(p_{\rm NL}^2)$, we will ignore it in computation of the theoretical parity-odd trispectrum below.

\subsection{Correlators}
\noindent Our ansatz for the primordial potential of \eqref{eq: parity-fourier} generates a parity-violating trispectrum that is the sum of four components, each of the form
\beq
    T^{1113}(\vk_1,\vk_2,\vk_3,\vk_4) &=& \av{\delta^{(1)}(\vk_1)\delta^{(1)}(\vk_2)\delta^{(1)}(\vk_3)\delta^{(3)}(\vk_4)}\\\nonumber
    &=& -ip_{\rm NL}\left[\vk_1\cdot\vk_2\times\vk_3\right] P_\phi(k_1)P_\phi(k_2)P_\phi(k_3)\left\{k_1^\alpha k_2^\beta k_3^\gamma-k_1^\alpha k_2^\gamma k_3^\beta + \text{(4 perms.)}\right\},
\eeq
summing over a total of six permutations. The full trispectrum becomes
\beq
    T_\phi(\vk_1,\vk_2,\vk_3,\vk_4) = -ip_{\rm NL}\left[\vk_1\cdot\vk_2\times\vk_3\right] P_\phi(k_1)P_\phi(k_2)P_\phi(k_3)k_1^\alpha k_2^\beta k_3^\gamma+\text{23 perms.} + \mathcal{O}(p_{\rm NL}^2),
\eeq
where odd permutations of $\{\vk_1,\vk_2,\vk_3,\vk_4\}$ pick up a negative sign due to the cross product. As discussed in \S\ref{sec: theory}, this is odd under exchange of $\vk\to-\vk$ (which leads to parity violation) and purely imaginary. Whilst this polynomial form is not the most general parity-odd trispectrum considered, it is likely that other physical models of parity violation (such as those considered in \citep{Philcox:2023ffy,Cabass:2022oap,Philcox:2022hkh}), can be projected onto templates such as the above, and thus efficiently constrained (analogous to the use of equilateral, orthogonal, and local shapes in bispectrum studies). 

It is further instructive to consider the other primordial correlators. First, there is strictly a parity-even contribution to the trispectrum: this appears only one-loop however, requiring diagrams of the form $T^{3311}$, which are suppressed by a factor of $p_{\rm NL}A_s$. In contrast, the primordial bispectrum is zero at all orders in $p_{\rm NL}A_s$; this occurs since $\phi^{\rm NG}$ contains only pieces linear and cubic in $\phi^{(1)}$, thus all Wick contractions contain an odd number of fields, and hence evaluate to zero. Finally, we note that the power spectrum na\"ively contains a parity-sensitive $13$-diagram, but this is equal to zero:
\beq
    P^{13}(\vk) = ip_{\rm NL}k^\alpha P_\phi(k)\int_{\vp}\left[\vk\cdot\vp\times-\vp\right] p^{\beta+\gamma}P_\phi(p)+\text{2 perms.} = 0.
\eeq
At two-loop order, there exists a $33$-diagram, given by
\beq
    P^{33}(\vk) &=& p^2_{\rm NL}\int_{\vp_1\vp_2\vp_3}\delD{\vk-\vp_1-\vp_2-\vp_3}\left[\vp_1\cdot\vp_2\times\vp_3\right]^2 P_\phi(p_1)P_\phi(p_2)P_\phi(p_3)\\\nonumber
    &&\,\times\,p_1^\alpha p_2^\beta p_3^\gamma\left\{p_1^\alpha p_2^\beta p_3^\gamma+p_1^\beta p_2^\gamma p_3^\alpha+p_1^\gamma p_2^\alpha p_3^\beta-p_1^\alpha p_2^\gamma p_3^\beta-p_1^\beta p_2^\alpha p_3^\gamma-p_1^\gamma p_2^\beta p_3^\alpha \right\}.
\eeq
This is generically suppressed by $\left(p_{\rm NL}A_s\right)^2$ compared to the tree-level $P_\phi$ prediction. One can show that the contribution scales as $k^2$ in the infrared ($k\to 0$) and, at most, as $k^{2\,{\rm max}\{\alpha,\beta,\gamma\}+2}P_\phi(k)$ in the ultraviolet ($k\to \infty$), thus our modification to the initial conditions is well defined.\footnote{This differs to \citep{Scoccimarro:2011pz}, who find large divergences in $P_{22}$ for some forms of initial condition generation. The difference here is that, due to the imposed condition $\alpha+\beta+\gamma=-3$, we do not have derivative operators acting on the entire $\phi^{(3)}$, which would yield reciprocal powers of $k$ in the infrared limit.} In practice, this contribution is removed via the rescaling of \eqref{eq: IC-rescaling}.

\subsection{Validation}\label{sec:validationICs}

\noindent To validate the initial conditions generation procedure we generate 1000 simulations of the primordial potential: 500 with $p_\mathrm{NL}=+10^6$ and 500 with $p_\mathrm{NL}=-10^6$. We then performed a range of tests including verifying that our rescaling, \eqref{eq: IC-rescaling}, removes the majority of the $p_\mathrm{NL}^2$ power spectrum contribution and verifying that we did not generate a spurious primordial bispectrum. A key test is to examine the primordial trispectrum, and compare its amplitude to the above theory model. Thus we applied estimators for the binned parity-even and parity-odd trispectrum, described in detail in Appendix \ref{app: estimator}, to test that we did not generate a parity-even trispectrum and that we generated the correct parity-odd signal. For both estimators we used six bins linearly spaced with bin centers spaced between $k=0.02\hMpc$ and $k=0.12\hMpc$. 

\begin{figure}
    \centering
    \includegraphics[width=0.78\textwidth]{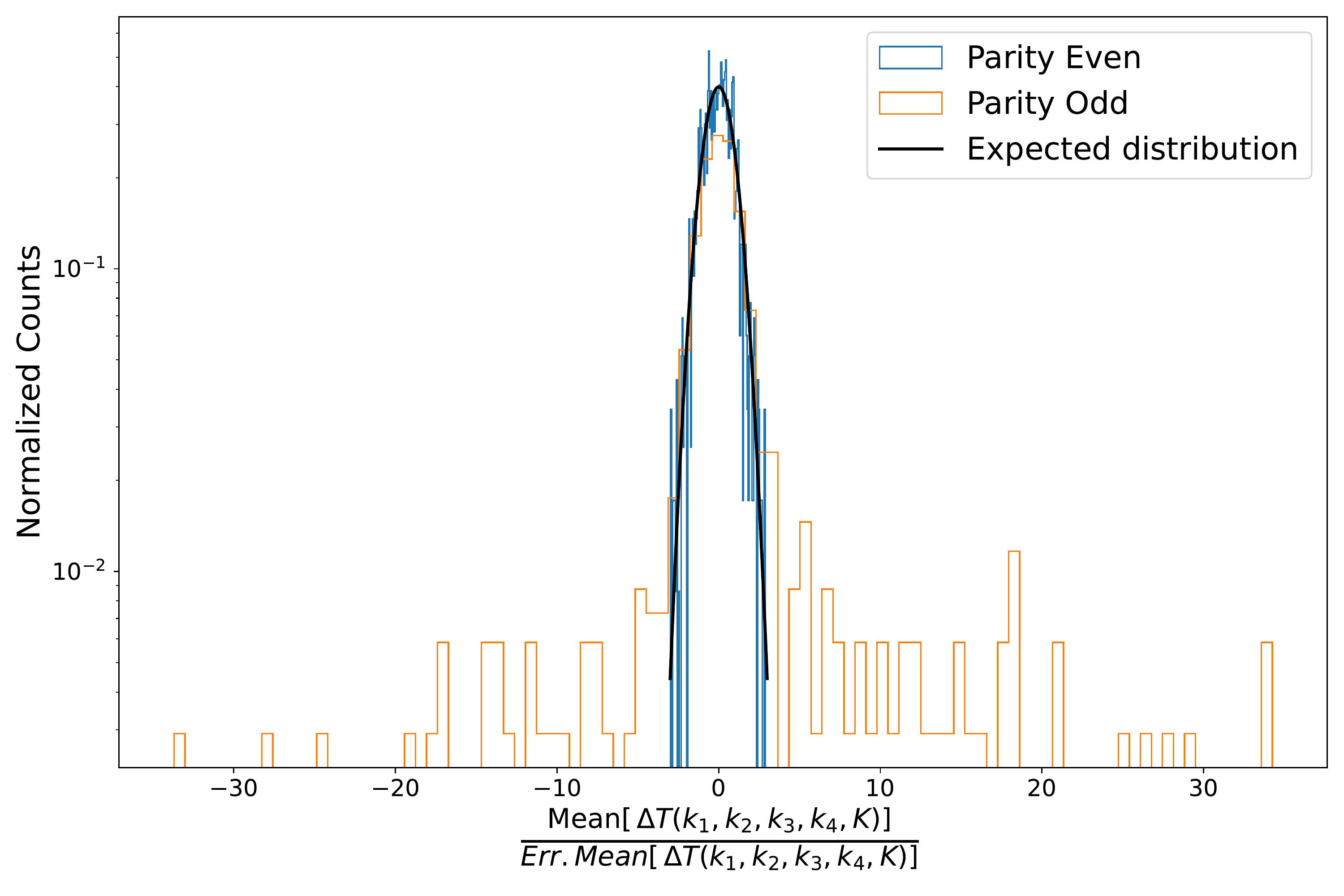}
    \caption{As a validation of the parity violating initial conditions, we measure the binned parity-even and parity-odd trispectrum signals from the primordial potential, cancelling sample variance by computing the difference between this measurement and a Gaussian simulation with matched phases. Here, we compare the mean trispectrum to the error on the mean, on a bin-by-bin basis (using $500$ simulations of each type). It can be seen that the parity-even results are consistent with noise, \textit{i.e.}\,no parity-even trispectrum is present, however there is clear evidence of a parity-odd trispectrum.}
    \label{fig: IC_hists}
\end{figure}

\begin{figure}
    \centering
    \includegraphics[width=0.78\textwidth]{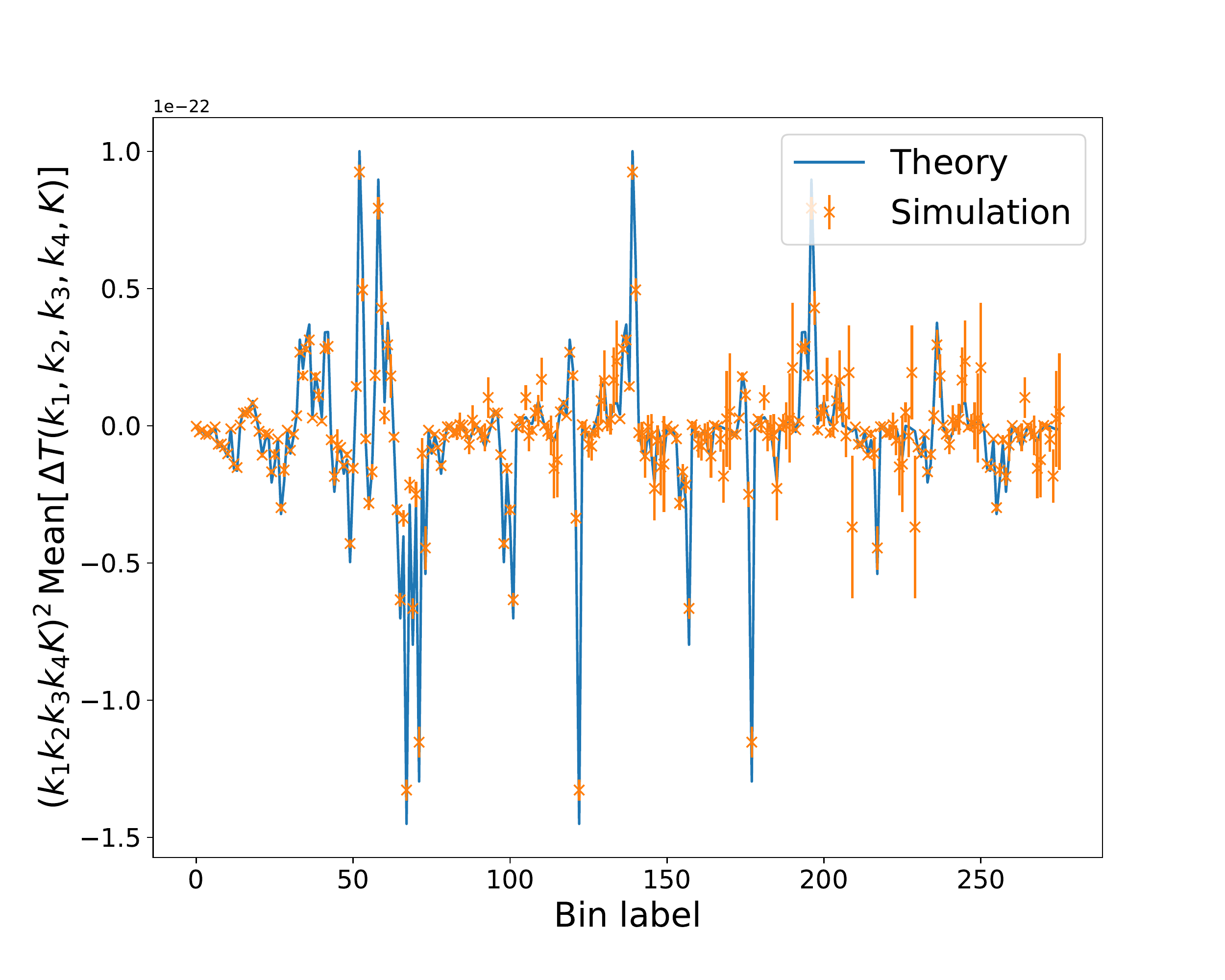}
    \caption{A comparison of the theoretical and simulated parity-odd trispectrum signals. To reduce noise we compute the difference between simulations with $p_\mathrm{NL}=10^6$ and Gaussian simulations with an identical random seed. Theoretical predictions are obtained as in Appendix \ref{app: tl-theory}, discretizing each bin into a number of sub-bins and averaging the corresponding trispectra. The high level of agreement provides a stringent test that our initial condition generation produces the correct signal.}
    \label{fig: IC_theory_sims}
\end{figure}

In Fig.~\ref{fig: IC_hists} we show the significance of the detection of the parity-even and parity-odd trispectra. No parity-even trispectrum is detected, as is expected, and there is strong evidence for a parity-odd trispectrum. To validate that the correct parity-odd trispectrum signal is generated, we compare the measured trispectrum to the theoretical expectation -- the computation of the expected binned signal is given in Appendix \ref{app:TheoryParityOddTris}. As can be seen in Fig.~\ref{fig: IC_theory_sims}, the theoretical trispectrum and measured trispectrum show great agreement across a range of scales and tetrahedron configurations.

\section{Parity-Violation in Practice: Simulations}\label{sec: sims}


\noindent Armed with the parity-violating initial conditions, we now proceed to run the forward model, \textit{i.e.}\ to generate parity-violating $N$-body simulations. These are generated in an identical manner to the \textsc{quijote} simulations \citep{Villaescusa-Navarro_2020,Coulton_2023}. We provide a brief summary of thie suite below and refer the reader to \citep{Villaescusa-Navarro_2020} and \url{https://quijote-simulations.readthedocs.io} for more details. Table \ref{tab:simulation_params} summarizes the key simulation parameters used in this work.

\begin{table*}
    \centering
    \begin{tabular}{c| c c c c c c c  c c c c c c}
 Type & $\Omega_m$ & $\Omega_\Lambda$  & $ \Omega_b $ & $ \sigma_8 $  & $h$ & $n_s$ & $ \sum m_\nu $  & $w$ & $L_{\rm box}$ &$ N_{\mathrm{particles}}^{\frac{1}{3}}$ &Realizations & $p_{\rm NL}$ & $M_{\rm min}$  \\ 
  & &  &  & & &  & (eV)  & & (Mpc/h) & & & & (M$_\odot$/h) \\
    \hline \hline 
 Parity$+$ & 0.3175 & 0.6825 & 0.049 & 0.834 & 0.6711 & 0.9624 & 0.0  & -1 &  1000 & 512 & 500 & $+10^6$
 & $6.56\times10^{11}$  \\
 Parity$-$ & 0.3175 & 0.6825 & 0.049 & 0.834 & 0.6711 & 0.9624 & 0.0  & -1 &  1000 & 512 & 500 & $-10^6$ 
 & $6.56\times10^{11}$  \\
   \end{tabular}
    \caption{Simulation parameters for the \textsc{quijote-odd} suite. These use the same parameters as the \textsc{quijote} and \textsc{quijote-png} simulations \citep{Villaescusa-Navarro_2020,Coulton_2023} , but feature non-zero parity-violation parameters $p_{\rm NL}$. In each case, the initial conditions are generated with a suitably modified version of the 2LPTIC code \citep{Scoccimarro:2011pz}.}\label{tab:simulation_params}
\end{table*}

First a realization of the primordial potential is generated on a $1024^3$ grid and the parity-violating trispectrum is added, via \eqref{eq: parity-transform}. In this work we generate simulations with $p_\mathrm{NL}=\pm 10^6$. These values were chosen as the size of the induced trispectrum is large enough to generate measurable signals, but small enough such that the higher order corrections ($\sim\mathcal{O}(p_\mathrm{NL}^2)$) are still small. We ran a small number of simulations with different $p_\mathrm{NL}$ values to validate this choice. Next, we perform a rescaling of the initial conditions to mitigate the impact of the $p_\mathrm{NL}^2$ terms using \eqref{eq: IC-rescaling}. The initial conditions are then evolved to $z=0$ using a transfer function computed by CAMB \citep{Lewis_2000} and then rescaled back to $z=127$ using a scale-independent growth factor. The density field at $z=127$ is then combined with 2LPT to compute particle displacements and peculiar velocities for glass-distributed particles.

The particles are then evolved using \textsc{gadget-3}, a treePM code \citep{Gadget}. In total we run 1,000 simulations: 500 with $p_{\rm NL}=+10^6$ and 500 with $p_{\rm NL}=-10^6$. All simulations contain $512^3$ particles in a comoving periodic box of $(1000\Mpch)^3$. The value of the parameters used to control the accuracy and precision of the different integrators are identical to those used for the \textsc{quijote} simulations \citep{Villaescusa-Navarro_2020}. Halos are identified using both the Friends-of-Friends (FoF) algorithm \citep{FoF} and \textsc{rockstar} \citep{Behroozi_2013}. We note that while FoF halo catalogs only include positions, masses, and velocities, the Rockstar catalogs include a much richer set of properties such as angular momentum and radius, which will be of great use in \S\ref{sec: results}. Finally, we create over-density grids from the output particle positions and halo catalogs. This uses the `Cloud-in-Cell' grid assignment scheme \citep{Hockney_1981} to produce grids of size $512^3$. 

\section{Parity-Violation in Practice: Late-Time Observables}\label{sec: results}

\noindent In this section we explore a range of commonly considered statistical probes and examine how they are changed by the presence of the parity-violating primordial trispectrum. Here, our focus is on the $z=0$ universe; similar results were found at $z=1$. Furtherore, to isolate the terms linear in $p_\mathrm{NL}$ and remove any $p_\mathrm{NL}^2$ terms, which are not parity violating, we show most of the results as the difference between the $p_\mathrm{NL}=10^6$ and $p_\mathrm{NL}=-10^6$ simulations. 

\begin{figure}
    \centering
    \includegraphics[width=0.48\textwidth]{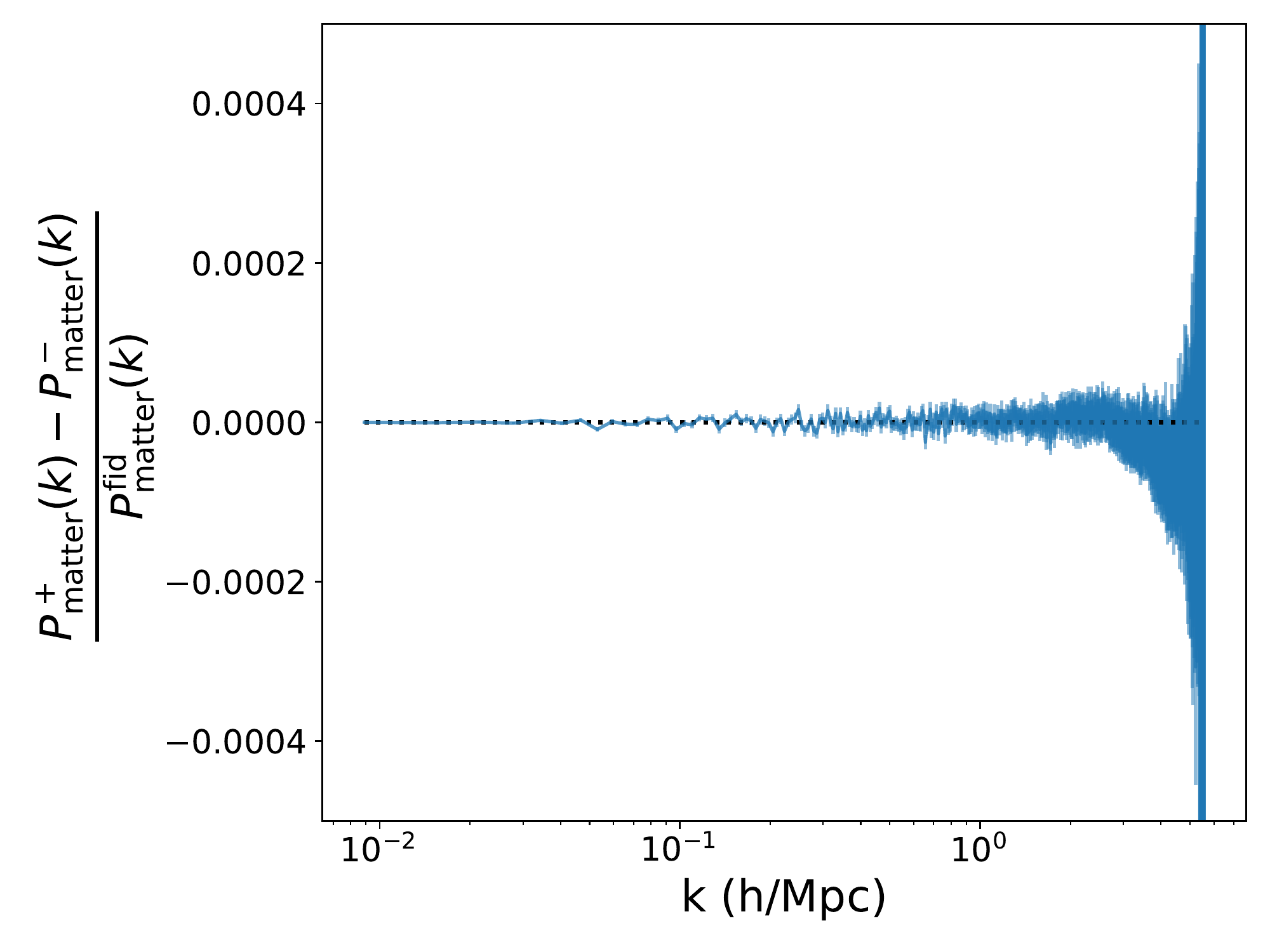}
    \includegraphics[width=0.48\textwidth]{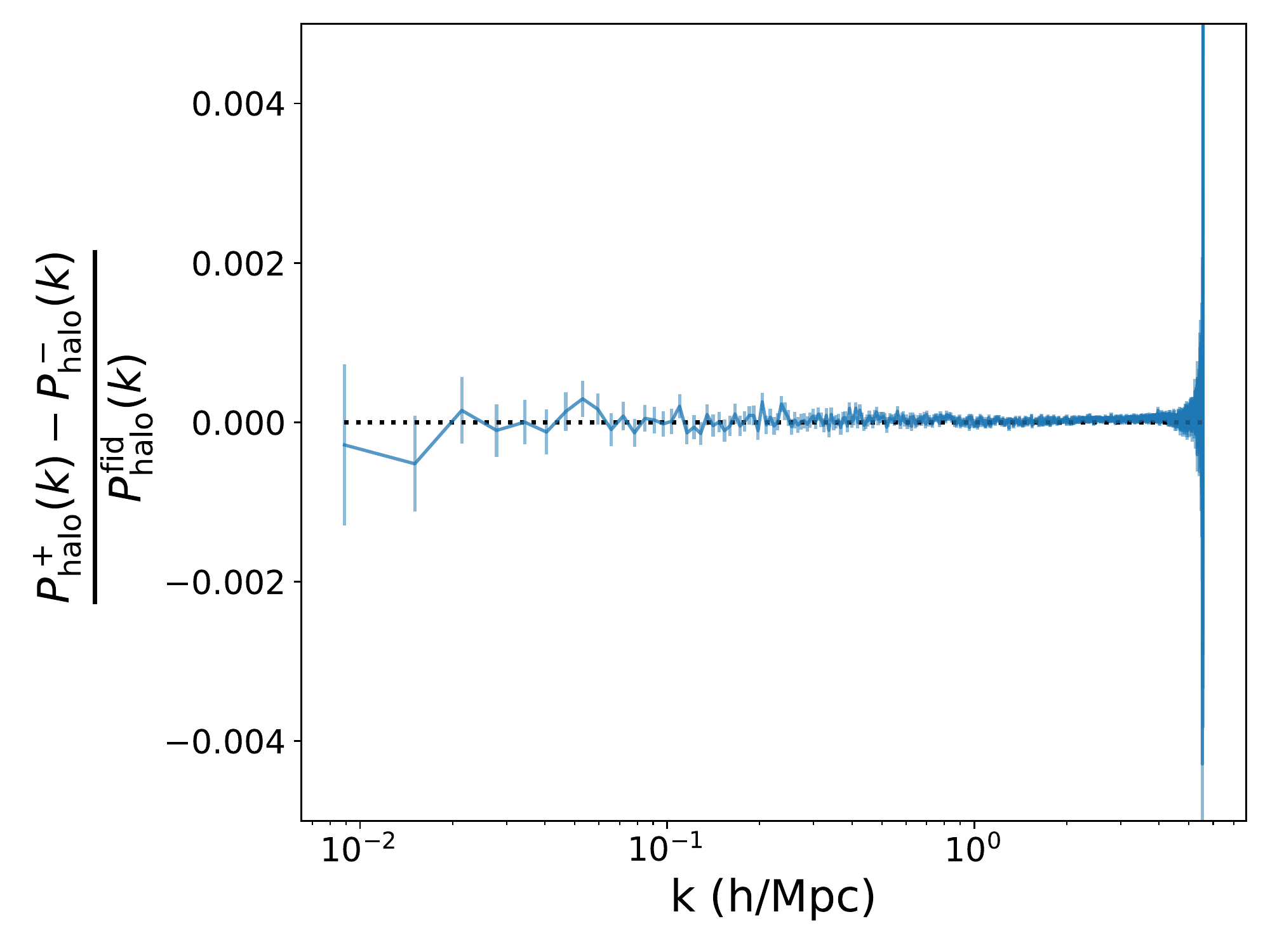}
    \caption{The response of the $z=0$ matter power spectrum, \textbf{left panel}, and the halo power spectrum, \textbf{right panel} to parity-violating non-Gaussianity. This is computed as the difference between 500 simulations with $p_\mathrm{NL}=10^6$ and $p_\mathrm{NL}=-10^6$. The halo power spectrum uses all halos with $M\geq 3.2\times 10^{13}$ M$_\odot$/h.  At leading order in $p_\mathrm{NL}$ both statistics are unchanged. The error bars on both plots denote the error on the mean, as measured with the 500 simulations.}
    \label{fig: pk_h_pk_m}
\end{figure}

\begin{figure}
    \centering
    \includegraphics[width=0.48\textwidth]{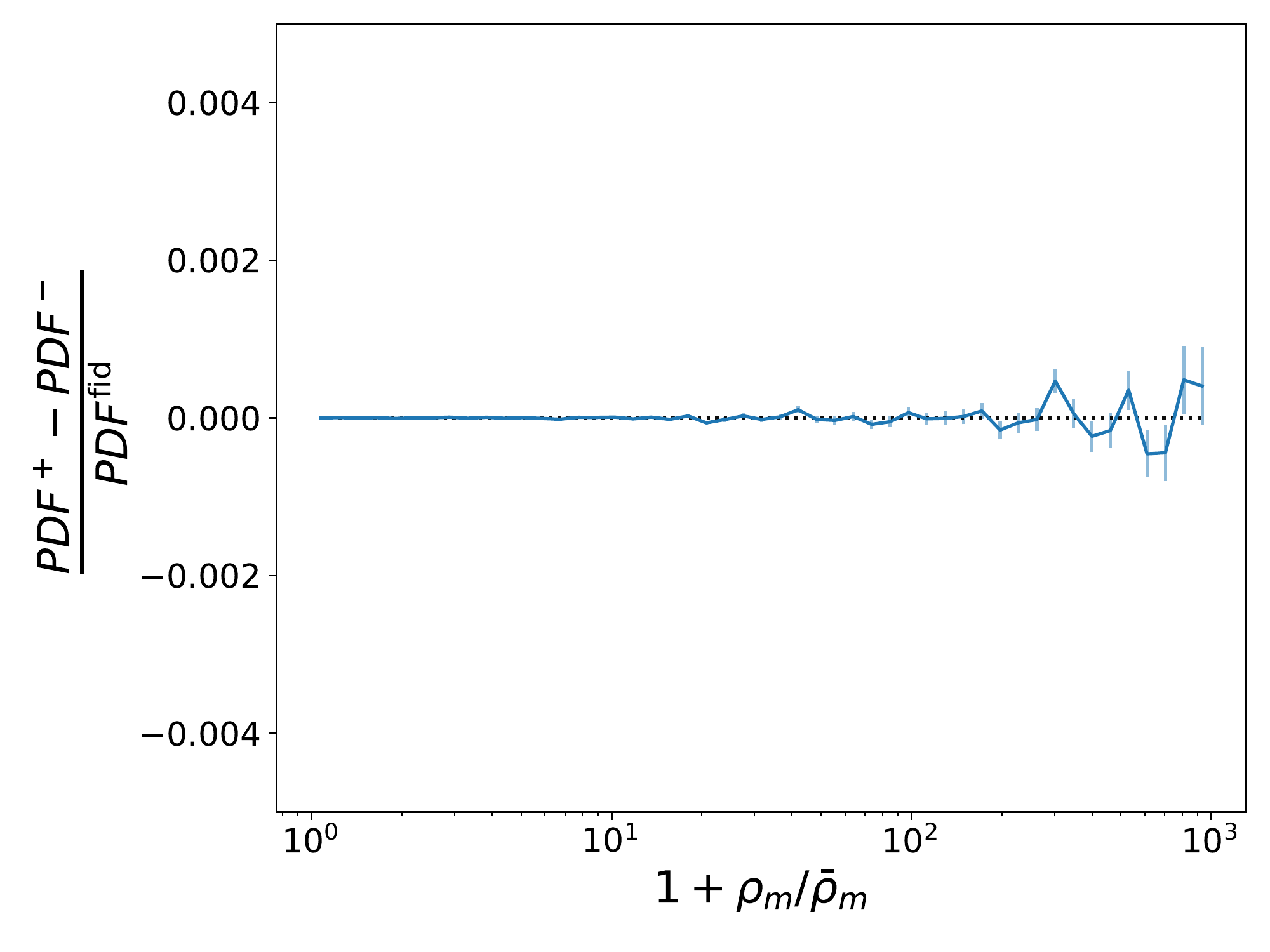}
    \includegraphics[width=0.48\textwidth]{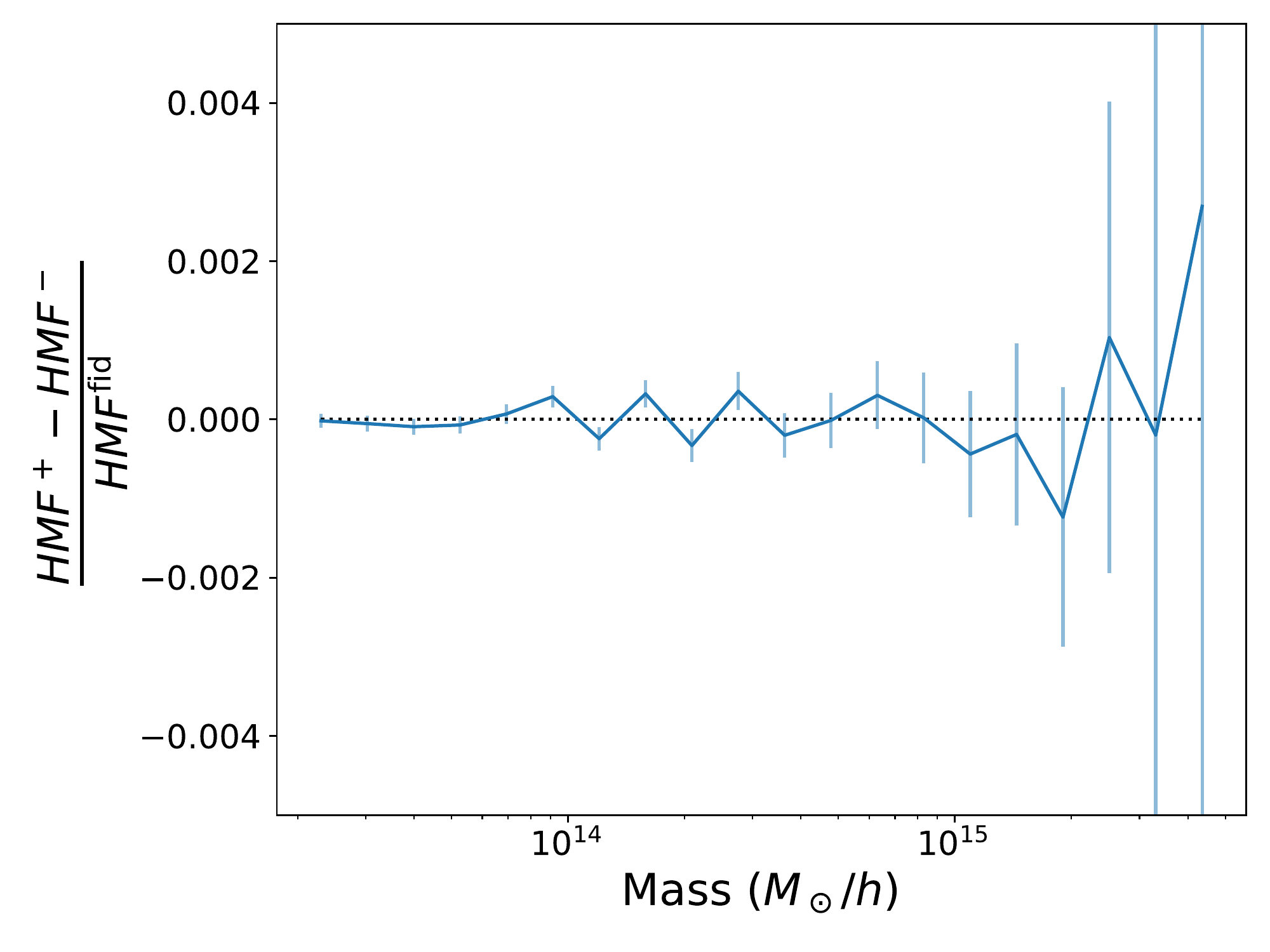}
    \caption{The response of the $z=0$ probability density function (PDF, \textbf{left panel}) and the halo mass function (HMF, \textbf{right panel}, computed using the difference of 500 simulations with $p_\mathrm{NL}=10^6$ and $p_\mathrm{NL}=-10^6$. Despite the presence of a large level of primordial non-Gaussianity, both statistics are unaffected. The error bars are the error on the mean. Note that we consider the PDF of 1+$\rho/\bar{\rho}$, where $\bar{\rho}$ is the mean density, to allow the logarithmic axis.}
    \label{fig: pdf_hmf}
\end{figure}

\begin{figure}
    \centering
    \includegraphics[width=0.48\textwidth]{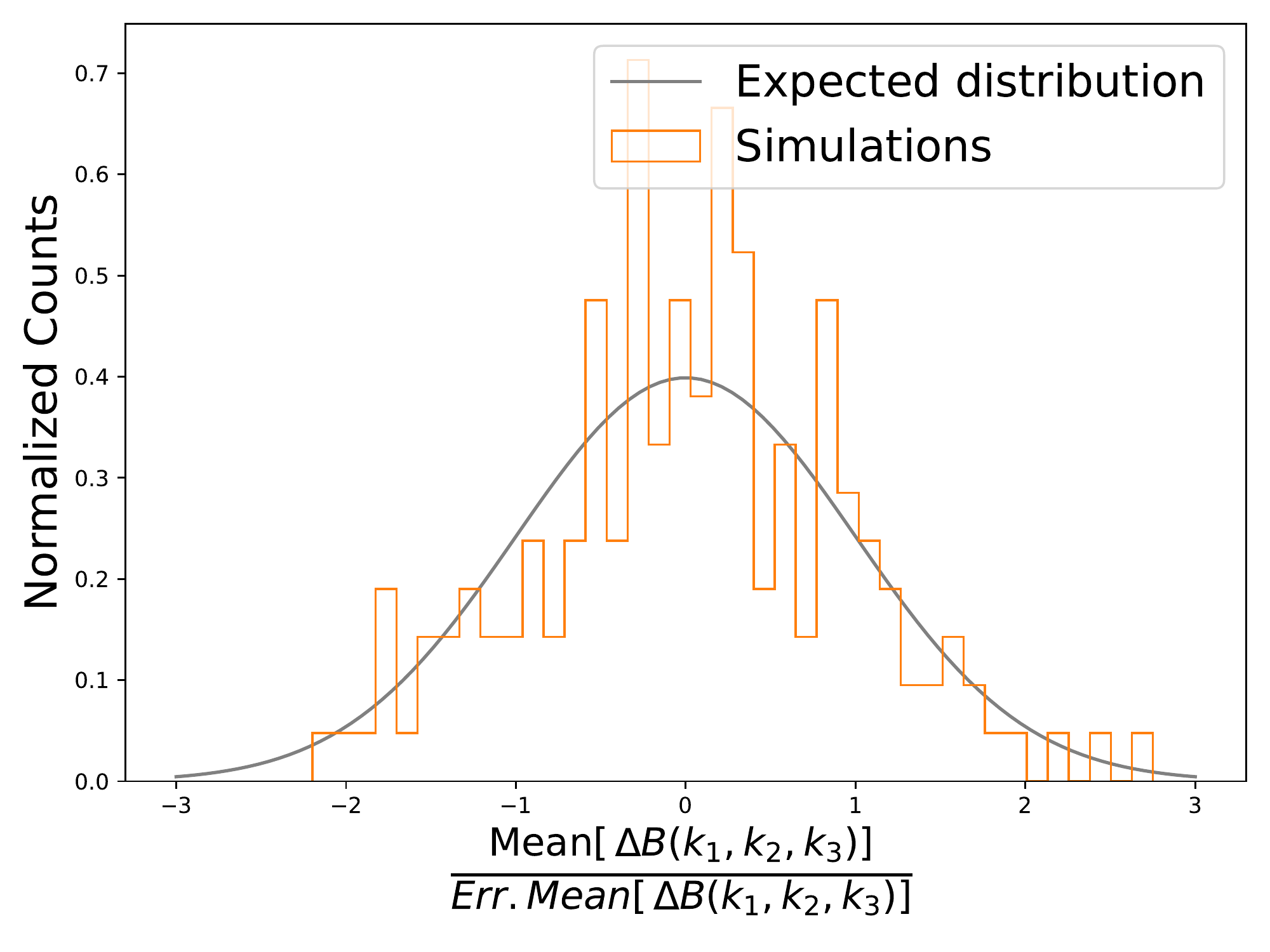}
    \includegraphics[width=0.48\textwidth]{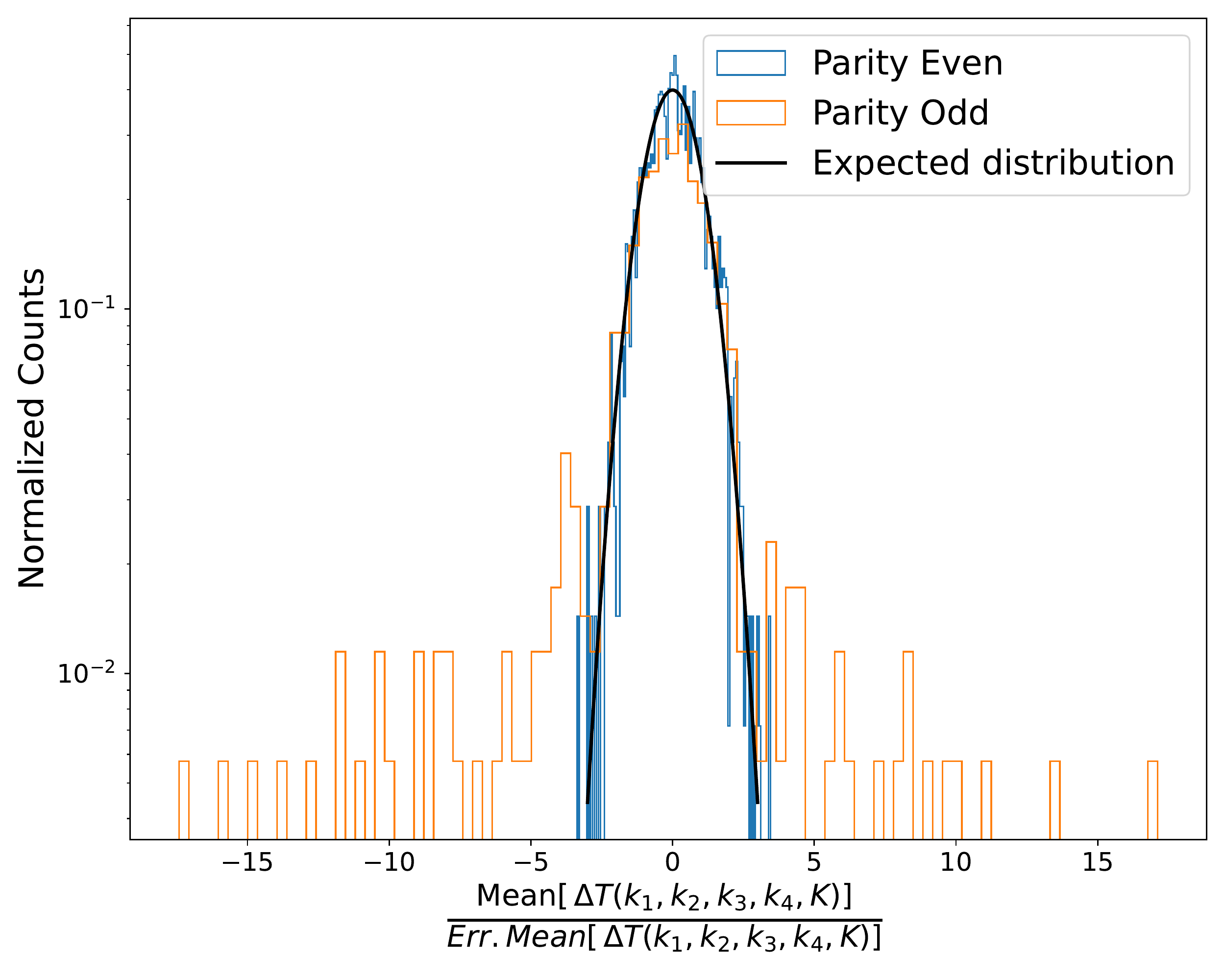}
    \caption{The impact of the parity-violating trispectrum on the $z=0$ matter bispectrum (\textbf{left panel)} and the parity-even and parity-odd matter trispectrum (\textbf{right panel}) normalized by the standard error on the mean for each bispectrum/trispectrum bin. As the distribution of bispectrum measurements is consistent with no signal (the blue line), we can conclude that no detectable bispectrum is sourced. The parity-even trispectrum shows a similar result -- no signal is detectable. However, we find strong evidence for a parity-odd trispectrum, implying that this quantity is preserved through cosmic time (though its amplitude is modulated by non-linear effects). Note that as we show the distribution of the difference between the $p_\mathrm{NL}=10^6$ and $p_\mathrm{NL}=-10^6$ simulations, we cancel most of the cosmic variance. }
    \label{fig: ispec_z=0_hists}
\end{figure}

\subsection{Power Spectrum}
\noindent We compute the power spectrum of the halo and matter field using \textsc{pylians} \citep{Pylians}, in a similar manner to \citep{Coulton_2023,Coulton_2023b}. In Fig.~\ref{fig: pk_h_pk_m} we show the impact of the parity violation PNG on the power spectrum: averaging over realizations, there is seen to be no effect in either the matter or halo power spectrum. Note that the matter power spectrum shows a small dip at small scales and similar effects were seen in the initial conditions.  This is not thought to be a signature of $p_\mathrm{NL}$ in the power spectrum.  This deviation arises as the quadratic (parity-even) terms in $p_\mathrm{NL}$ lead to strong correlations between $P(k)$ and $P(k')$ and an increase in the small scale power spectrum variance. This effect will thus average down with a very large number of simulations. In smaller ensembles, this can give rise to features that appear significant as many highly correlated bins can deviate from zero. In our case, the deviation of any single bin is $\sim1\sigma$ (in the mean) and an examination of the correlation matrix shows these bins are strongly correlated. 

\subsection{Matter Probability Density Function and the Halo Mass Function}

\noindent In Fig.~\ref{fig: pdf_hmf} we explore how the matter probability density function (PDF) and halo mass function (HMF) are affected. The PDF is computed from the matter density grids, as discussed in Section \ref{sec: sims}, whilst the halo mass function is computed from the FoF halo catalogs as in \citep{Jung_2023}. As can be seen, we find no detectable average signal of $p_\mathrm{NL}$ in either statistic. This agrees with the conclusions of \S\ref{sec: theory}.

\subsection{Bispectrum}
\noindent Next, we compute the binned bispectrum in a similar manner to \citep{Foreman_2020,Coulton_2023,Coulton_2023b}, using 10 bins linearly spaced between $k=0.013\hMpc$ to $k=0.19\hMpc$.  In the left panel of Fig.~\ref{fig: ispec_z=0_hists}, we show the ratio of the difference in bispectrum bin value between 500 $p_\mathrm{NL}=10^6$ and $p_\mathrm{NL}=-10^6$ simulations to the error on the mean of that bin. Once again, we find no evidence of an induced bispectrum.

\subsection{Trispectrum}

\noindent Using the estimators in Appendix \ref{app: estimator}, we measure the parity-even and parity-odd trispectrum with the same binning scheme as for the primordial trispectrum in Section \ref{sec:validationICs}. The parity-even trispectrum measurements are consistent with no induced trispectrum signal, as expected. In contrast, the parity-odd trispectrum shows strong evidence of a signal. Note that this is computed from the difference of simulations with $p_\mathrm{NL}=10^6$ and $p_\mathrm{NL}=-10^6$, which cancels most of the cosmic variance. Without this technique, the signal-to-noise on the parity-odd signal would be dramatically reduced. In its presence, we find large detection significances in individual bins, up to $40\sigma$ in the mean.

\subsection{Halo Angular Momenta}
\noindent As discussed in \S\ref{subsec: theory-spin}, the angular momenta of simulated tracers can be parity-sensitive. Here, we consider two associated tests: (a) correlating halo angular momentum, $\mathbf{J}$, with the velocity field, (b) correlating halo angular momentum with a chiral proxy derived from the initial conditions. In each case, we use the angular momenta measured from the \textsc{rockstar} code \citep{Behroozi:2011ju}, dropping any subhalos, and considering a minimum mass cut of $3\times 10^{13}h^{-1}M_\odot$.

\begin{figure}
    \centering
    \includegraphics[width=0.48\textwidth]{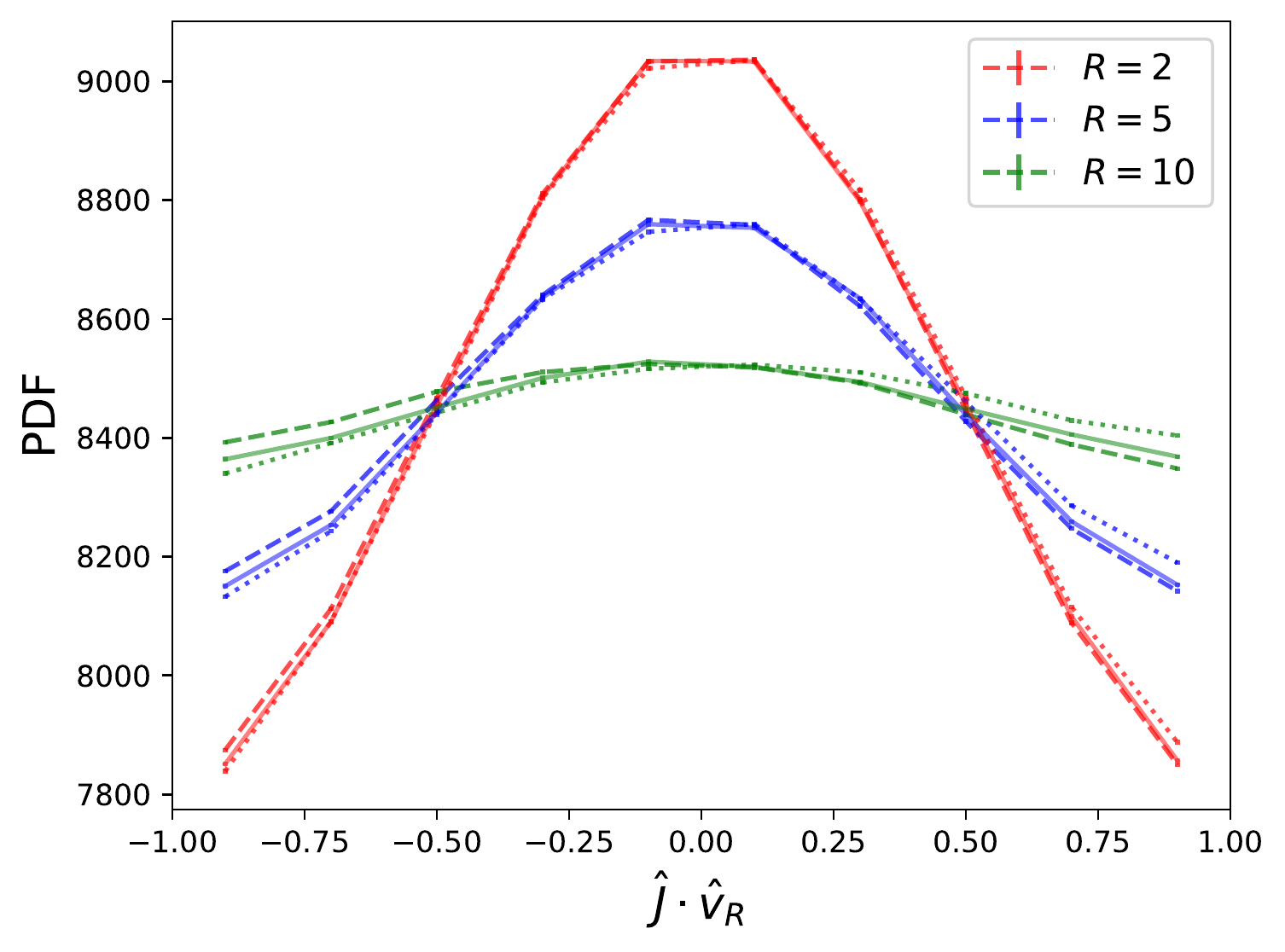}
    \includegraphics[width=0.48\textwidth]{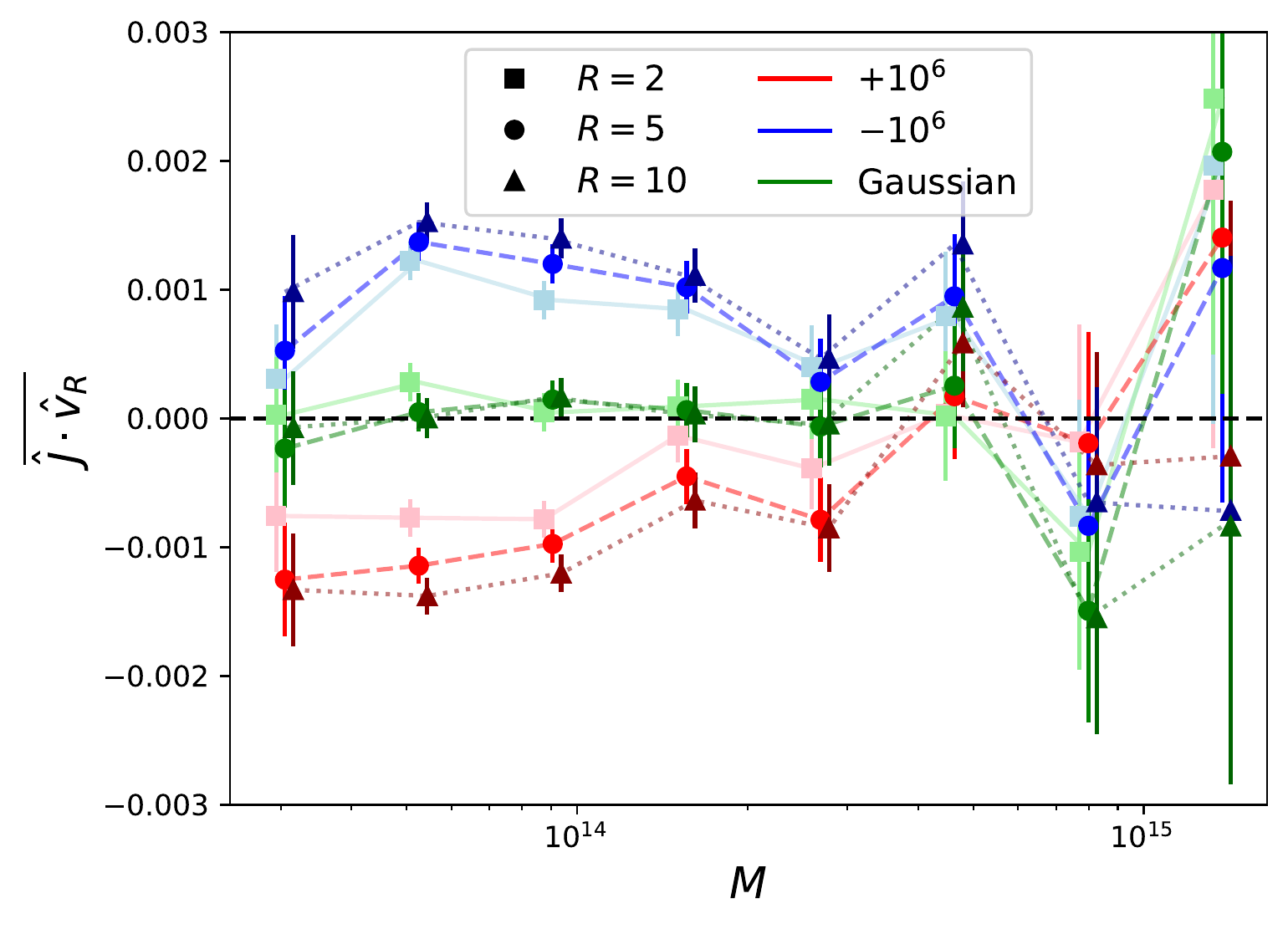}
    \caption{Correlation between redshift zero halo angular momentum, $\vec J$, and the velocity field, $\vec v_R$, measured from the initial conditions, smoothed on some scale $R$. \textbf{Left panel}: PDF of $\hat{\vec J}\cdot\hat{\vec v}_R$ from all halos in the three suites of simulations with $M>3\times 10^{13}h^{-1}M_\odot$. Results are shown for Gaussian simulations (solid line), $p_{\rm NL}=10^6$ (dashed lines) and $p_{\rm NL}=-10^6$ (dotted lines). The Gaussian distribution is symmetric, but we find asymmetry for the parity-breaking simulations, with some dependence on the smoothing scale $R$. \textbf{Right panel}: mean correlation averaged across eight mass bins for each suite of 500 simulations. We observe a clear non-zero signal proportional to $-p_{\rm NL}$, particularly at low masses.}
    \label{fig: spin-vel-correlation}
\end{figure}

In the first case, we compute the velocity field, $\vec v_R(\vr)$, from the $z=127$ snapshot (as $\vec v(\vk,z) = (-i\vk/k^2)f(z)\mathcal{H}(z)\delta(\vk,z)$ for growth rate $f(z)$ and conformal Hubble parameter $\mathcal{H}(z)$), smoothing on a scale $R\in\{2,5,10\}\Mpch$. The resulting field is interpolated to the halo positions, then the cosine $\hat{\vec J}\cdot\hat{\vec v}_R$ is computed for each halo in the dataset (dropping the moduli of $\vec J$ and $\vec v_R$, since they are scalar quantities, and thus parity insensitive). Fig.\,\ref{fig: spin-vel-correlation} shows the resulting distribution of $\hat{\vec J}\cdot\hat{\vec v}_R$ at redshift zero, averaged over $500$ simulations with $p_{\rm NL}\in\{0,\pm 10^6\}$. We find a distinctly non-uniform distribution for small smoothing scales (close to the Lagrangian radii of low mass halos), indicating that halo angular momenta correlate with the primordial density field. The Gaussian PDF appears symmetric; when $p_{\rm NL}$ is non-zero, we see asymmetries, which are characteristic signatures of parity-violation. This is clearly seen in the binned $\hat{\vec J}\cdot\hat{\vec v}_R$ statistic, which is expected to be (statistically) zero in the absence of parity violation. For $M\lesssim 3\times 10^{14}h^{-1}M_\odot$, we find a highly significant signal proportional to $-p_{\rm NL}$, indicating that the angular-momentum-velocity cosine is a good probe of parity-violation, matching the conclusion of \S\ref{sec: theory}. 

\begin{figure}
    \centering
    \includegraphics[width=0.48\textwidth]{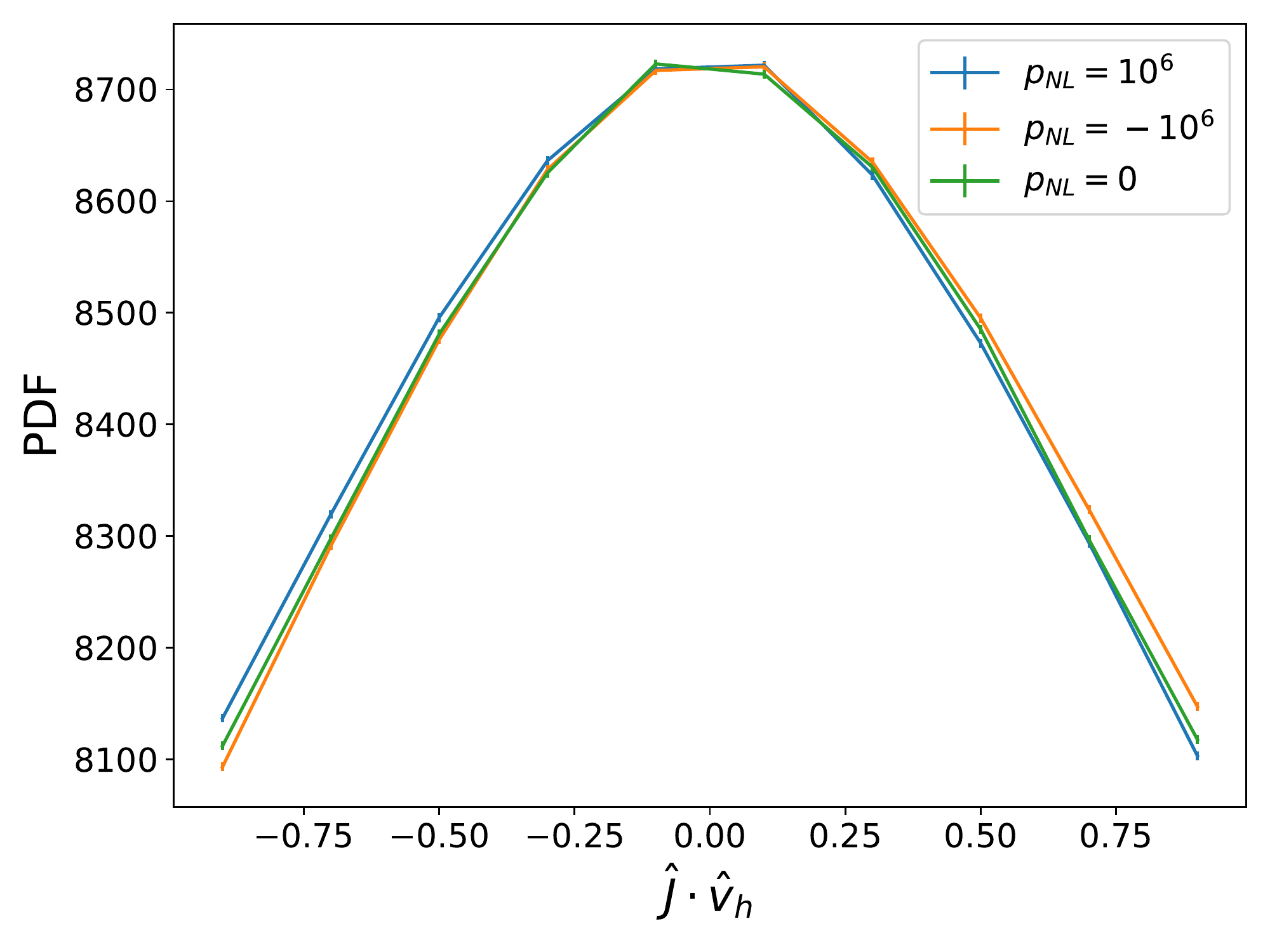}
    \includegraphics[width=0.48\textwidth]{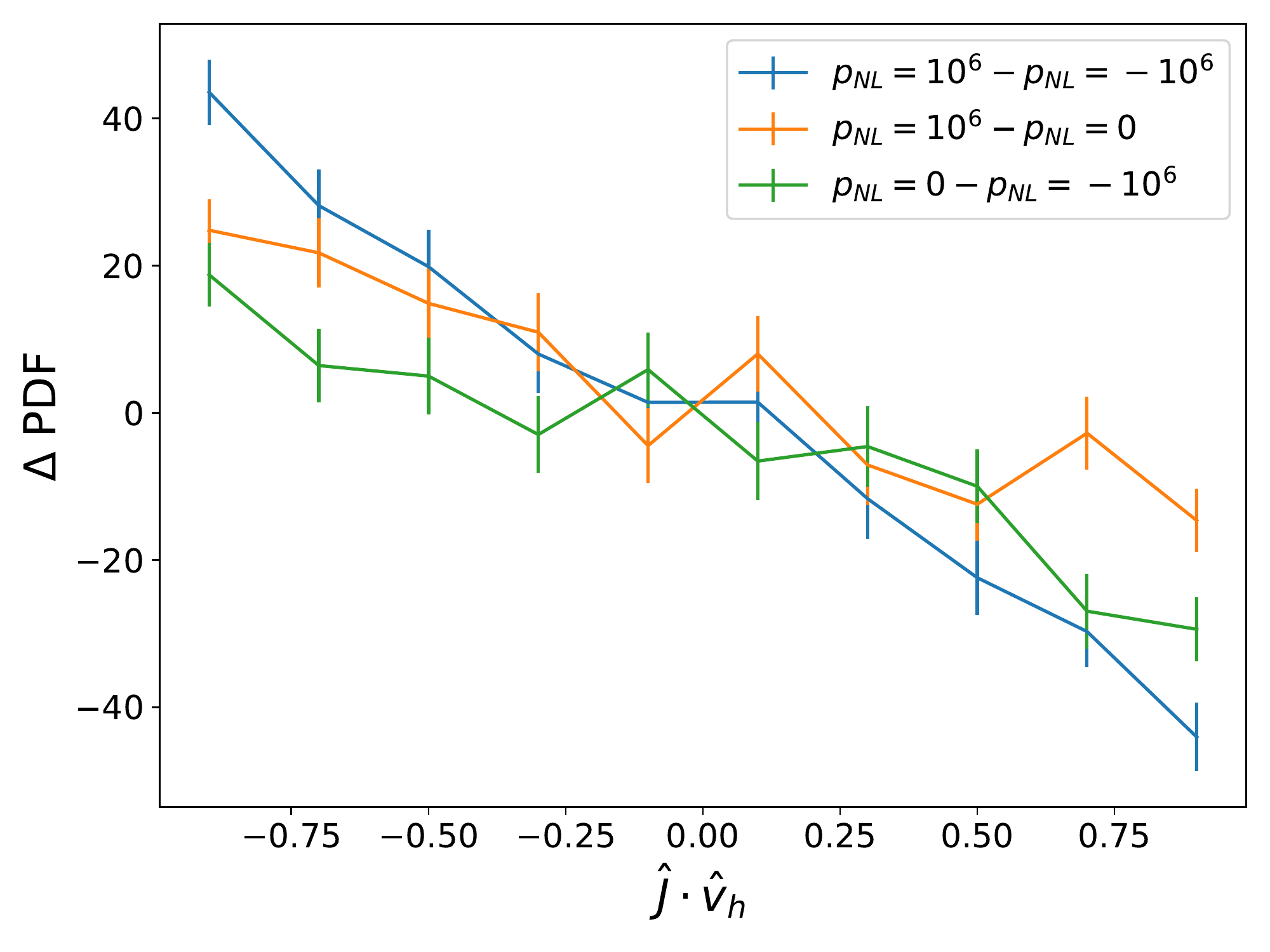}
    \caption{Correlation between redshift zero halo angular momentum, $\vec J$, and the halo velocity, $\vec{v}_h$, at $z=0$. As in Fig.\,\ref{fig: spin-vel-correlation}, we plot the PDF of all the \textsc{rockstar} halos (\textbf{left panel}) and the differential change (\textbf{right panel}). Whilst noisier than the equivalent correlation with the initial condition velocity field, this demonstrates that the parity violating signals can be seen using late-time measurements only. }
    \label{fig: spin-vel-correlation-lateTime}
\end{figure}

An alternative approach is to construct the $\hat{\vec J}\cdot\hat{\vec v}$ statistic from halo velocities measured by \textsc{rockstar}, removing the need for the initial conditions.\footnote{An alternative approach is to perform some flavor of reconstruction to obtain the velocity field, such as via BAO reconstruction techniques (using the continuity equation) or fully Bayesian methods \citep[cf.,][]{Motloch:2021lsw}.} The results for this approach are shown in Fig.~\ref{fig: spin-vel-correlation-lateTime}. This measurement shows reduced signal-to-noise, which we attribute to the contamination of the velocity effect by uncorrelated effects such as the Fingers-of-God contributions. It is expected that measurements using galaxy velocities, closer still to direct observations, would also be a clean probe for parity-violating signals. However, galaxies are expected to exhibit even larger noise. To examine this, we analyzed galaxies from the \textsc{camels} \citep{Villaescusa-Navarro_2021} simulation suite to support these statements. Unfortunately, the small size of these simulations, with boxes of side length $25\Mpch$, prevents a quantitative analysis. Qualitatively, variations of the cosmological parameters or the type of subgrid model in the \textsc{camels} suite led to symmetric changes in the $\hat{J}\cdot\hat{v}_h$ distribution and so cannot mimic the parity-violating signal. 

\begin{figure}
    \centering
    \includegraphics[width=0.48\textwidth]{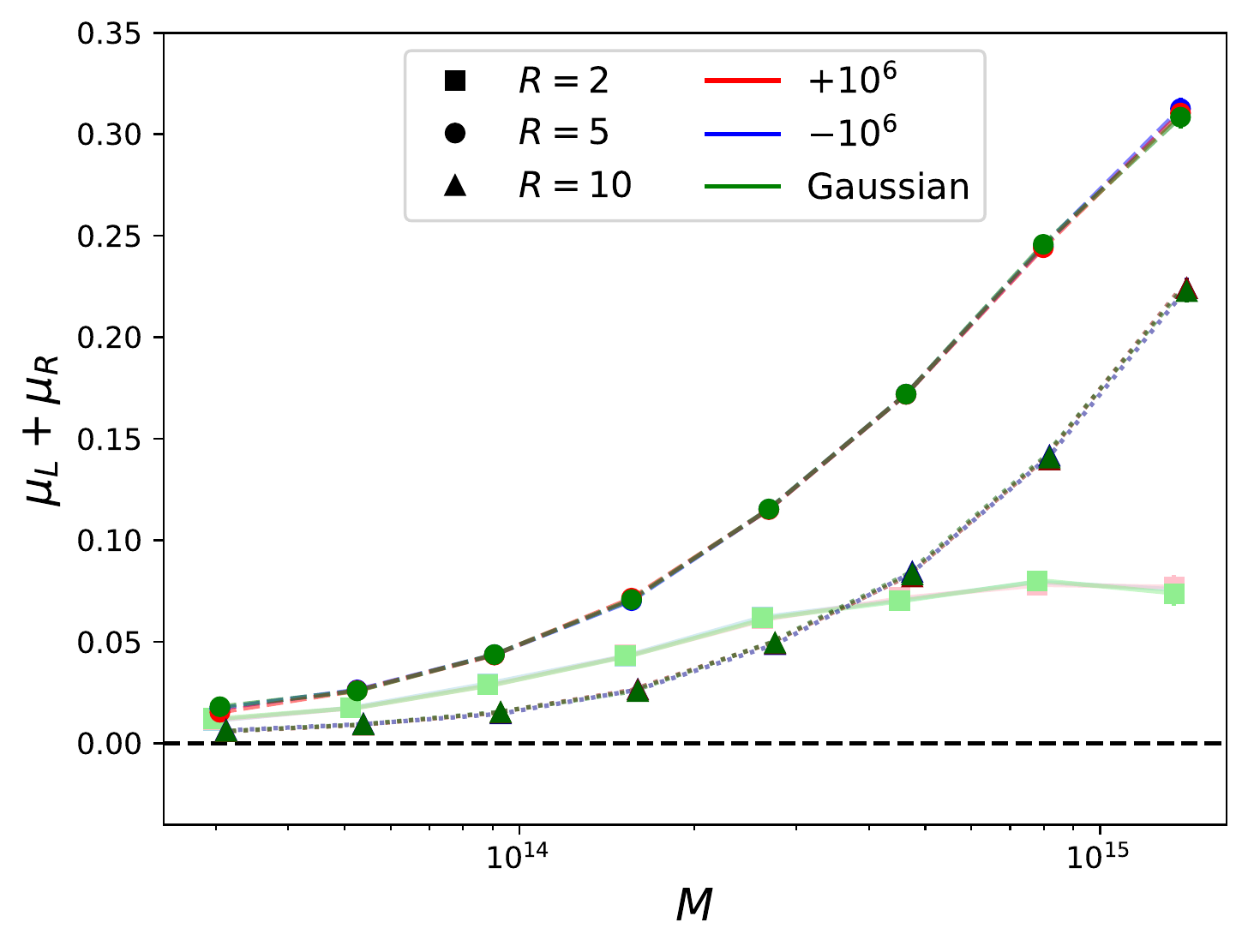}
    \includegraphics[width=0.48\textwidth]{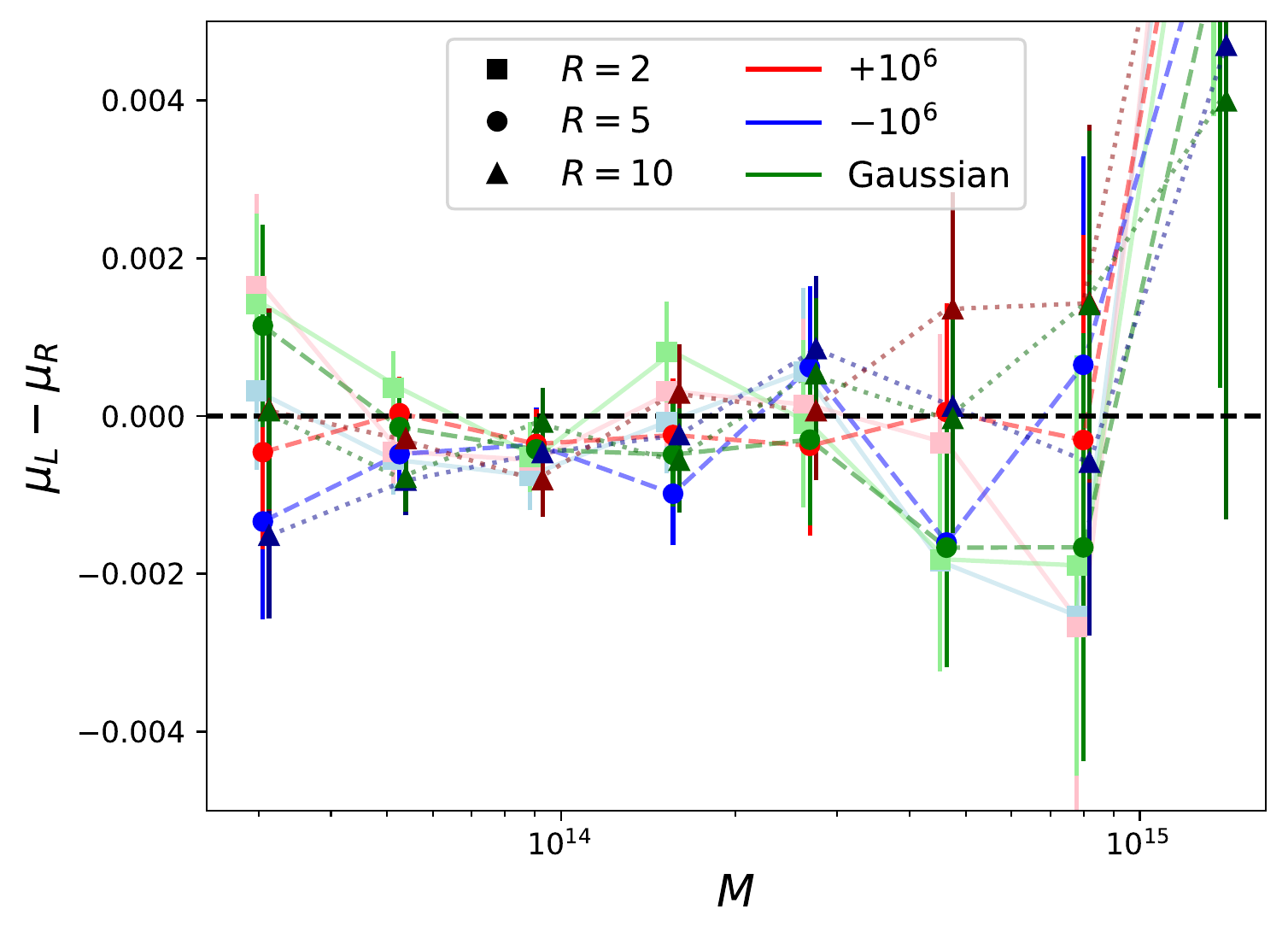}
    \caption{Cosine between redshift zero halo angular momentum, $\vec J$, and the helical templates, $\vec J_{L,R}$ defined from the initial density field, smoothed on some scale $R$, with $\mu_{L,R}\equiv \hat{\vec J}\cdot\hat{\vec J}_{L,R}$. \textbf{Left panel}: Parity-conserving contribution, $\mu_L+\mu_R$, binned in mass and averaged over 500 simulations. We find a strong correlation, matching the results of \citep{Yu:2019bsd}; this is not seen to depend on $p_{\rm NL}$. \textbf{Right panel}: Parity-violating contribution, $\mu_L-\mu_R$, binned as before. We find no discernible signal in this case, regardless of $p_{\rm NL}$, even though it is not nulled by symmetry arguments.}
    \label{fig: spin-spin-correlation}
\end{figure}

For a second probe, we correlate the halo angular momentum $\vec J$, with the $\vec J_{L,R}$ proxy defined from the smoothed initial potential $\phi$ and density $\delta$, as discussed in \citep{Yu:2019bsd,Motloch:2020qvx,Motloch:2021lsw,Motloch:2021mfz}. This is given by
\beq
    J_i^{\rm theory}(\vx) = \epsilon_{ijk}\partial^j\partial^m\phi_R(\vx)\partial^k\partial_m\delta_R(\vx), \qquad J_{L,R}^i(\vk) = \frac{1}{2}\left[\left(\delta_{\rm K}^{ij}-\hat{\vec k}^i\hat{\vec k}^j\right)\pm i\epsilon^{ijk}\hat{\vec k}_k\right]J^{\rm theory}_j(\vk)
\eeq
explicitly projecting onto the helical basis in the second equation. In Fig.\,\ref{fig: spin-spin-correlation}, we plot the cosines $\mu_{L,R}\equiv \hat{\vec J}\cdot\hat{\vec J}_{L,R}$ for each set of simulations, and a variety of smoothing scales. The parity-conserving contributions show a clear signal, echoing the conclusion of \citep{Yu:2019bsd}; the angular momenta of halos correlates with the primordial density field. Here, we observe largest correlations for smoothing scales $R=5\Mpch$, which is somewhat larger than that suggested in \citep{Yu:2019bsd}, since the halos in our simulations are more massive and have larger Lagrangian radii. We do not observe any differences between the two sets of simulations, indicating that any contributions quadratic in $p_{\rm NL}^2$ are small. Considering $\mu_L-\mu_R$, we find no evidence for parity-violation sourcing a helical angular momentum, with results consistent with zero for all values of $p_{\rm NL}$, mass bins, and smoothing. This result is not obvious \textit{a priori}, since $\mu_L-\mu_R$ could contain terms linear in $p_{\rm NL}$ (cf.\,\S\ref{subsec: theory-spin}). We conclude that any signatures are too small to see in our choice of initial conditions, likely due to the strong scale-dependence of the assumed primordial correlator. 

\section{Conclusion}\label{sec: conc}
\noindent By probing the parity-properties of the late-time Universe, we can place constraints on non-standard physics occurring during inflation or low-redshift structure growth. In this work, we have performed the first numerical study of mirror asymmetries in large-scale structure by generating and analyzing simulations with a particular form of parity-violating initial conditions. This corresponds to injecting an imaginary inflationary four-point function, proportional to an amplitude $ip_{\rm NL}$, and computing various statistics on the evolved simulations at low redshift.

Our principal conclusion is that early-Universe parity violation is difficult to detect in the late-Universe. Due to homogeneity and isotropy, the realization-averaged power spectrum, bispectrum, and halo mass function are not affected by our modifications to the initial conditions, even on arbitrarily non-linear scales and including redshift-space distortions. This has been explicitly verified with simulations, carefully accounting for parity-conserving $\mathcal{O}(p_{\rm NL}^2)$ contributions. To measure parity-violation from scalar observables, we must look to four-point functions or beyond, such as those used in the large scale structure analyses of \citep{Philcox:2022hkh,Hou:2022wfj}, following the methodology of \citep{2021arXiv211012004C}, and, the CMB analyses of \citep{Philcox:2023ffy}. Whilst non-linear evolution modifies the shapes of late-time trispectra, it does not source other spectra, and we find a clear detection of a late-Universe parity-violation using a novel trispectrum estimator.

A conceptually simpler test for parity-violation comes from tensorial quantities, such as the angular momentum of halos. Theoretically, parity-violation can be probed using both the correlation of halo angular momenta with velocity fields (a naturally occurring pseudo-scalar), and helical proxies extracted from the initial conditions \citep{Yu:2019bsd}. In our set-up, we find a strong correlation of angular momentum and velocity, scaling linearly with $p_{\rm NL}$, but no signal in the latter observable (though this is allowed theoretically).  Whilst such quantities can be straightforwardly extracted from $N$-body simulations, the realistic measurement is a little more nuanced, since we usually have access only to projected velocity fields (themselves contaminated with non-linear effects such as the Fingers-of-God effect), and there are non-trivial baryonic effects at play. These complications have, to some extent, already been solved in \citep{Motloch:2020qvx,Motloch:2021lsw,Motloch:2021mfz,Yu:2019bsd}, for example using reconstructions of the local density field, making the underlying statistic a potentially promising observable for future study. 

Finally, we make publicly available our simulation suite, \textsc{quijote-odd}, together with different data products such as halo catalogs and power spectra, to allow future exploration of simulations. A description of the data products and how to access them is available at \url{https://quijote-simulations.readthedocs.io/en/latest/odd.html} . We envisage that a number of other tests can be explored, for example, machine learning determinations of the subtle signatures of inflationary parity-violation. Another intriguing possibility is that our primordial injection could impact galaxy intrinsic alignments, due to the correlation between galaxy shapes and local properties such as angular momenta. Though one would require hydrodynamic simulations for a full study, this could potentially lead to unexplored signals in the $EB$ cross-correlation of galaxy shear.

\acknowledgments

\noindent We thank Ue-Li Pen and  Jiamin Hou for enlightening conversations, and are grateful to Eiichiro Komatsu and David Spergel for comments on the manuscript. All authors thank the Simons Foundation for support. This work was partly conceptualized at the ``Primordial Physics with Spectroscopic Surveys'' workshop in San Diego, hosted by Dan Green; the remainder stems from a bagel-fuelled argument between WRC and OHEP. 
\appendix

\section{Trispectrum Estimators}\label{app: estimator}

Following the discussion in \S\ref{sec: theory}, a general trispectrum of some real field $\delta$ can be written as
\beq\label{eq: Tl-general}
    \av{\delta(\vk_1)\cdots\delta(\vk_4)} = \delD{\vk_1+\vk_2+\vk_3+\vk_4}\left[\tau_+(k_1,k_2,k_3,k_4,K,K')+i(\vk_1\cdot\vk_2\times\vk_3)\tau_-(k_1,k_2,k_3,k_4,K,K')\right],
\eeq
where $\tau_{\pm}$ give parity-even and parity-odd components, and we have factorized out the scalar triple product in the latter. Each component depends on four side lengths ($k_i$) and two diagonals ($K\equiv |\vk_1+\vk_2|$ and $K'\equiv|\vk_1+\vk_3|$); in practice, it is necessary to parametrize by only one diagonal, else the trispectrum estimators are not separable. In the below, we consider how to estimate each component, and, in Appendix \ref{app: tl-theory}, relate this to theoretical trispectra. The parity-even estimator is additionally described in \citep{Jung:2023kjh}.

\subsection{Parity-Even Estimator}
Considering bins $q_i$ in $\vk_i$ and $Q$ in $\vK$, a general estimator for the parity-even component is given by:
\beq
    \widehat{T}_+(q_1,q_2,q_3,q_4,Q) \propto \int_{\vk_i\in q_i,\,\vK\in Q}\left[\prod_{i=1}^4\delta(\vk_i)\right]\delD{\vk_1+\vk_2-\vK}\delD{\vk_3+\vk_4+\vK},
\eeq
where we explicitly integrate over the internal momentum $\vK$ and drop a normalization factor. This can be efficiently implemented by rewriting the Dirac deltas as exponential integrals, yielding
\beq
    \widehat{T}_+(q_1,q_2,q_3,q_4,Q) &\propto& \int_{\vK\in Q}\left[\int d\vx\,\left(\int_{\vk_1\in q_1}\delta(\vk_1)e^{-i\vk_1\cdot\vx}\right)\left(\int_{\vk_2\in q_2}\delta(\vk_2)e^{-i\vk_2\cdot\vx}\right)e^{i\vK\cdot\vx}\right]\\\nonumber
    &&\,\times\,\left[\int d\vy\,\left(\int_{\vk_3\in q_3}\delta(\vk_3)e^{-i\vk_3\cdot\vy}\right)\left(\int_{\vk_4\in q_4}\delta(\vk_4)e^{-i\vk_4\cdot\vy}\right)e^{-i\vK\cdot\vy}\right].
\eeq
Practical computation is achieved via repeated Fourier transforms, first computing the $N_k$ transforms of $\delta$ (for $N_k$ bins), then assembling the $N_k(N_k+1)/2$ pairs (from the $\vx,\vy$ integrals), and lastly summing over $\vK$ for all $Q$ bins of interest. The typical normalization includes the same functions but with the $\delta(\vk)$ fields replaced with unity, such that the normalization counts the total number of tetrahedra in a given bin. 

The above estimator has the following symmetries:
\beq
    \widehat{T}_+(q_1,q_2,q_3,q_4,Q) = \widehat{T}_+(q_2,q_1,q_3,q_4,Q) = \widehat{T}_+(q_3,q_4,q_1,q_2,Q);
\eeq
in addition, modes with $\widehat{T}_+(q_1,q_3,q_2,q_4,Q)$ are partially covariant with $\widehat{T}_+(q_1,q_2,q_3,q_4,Q)$, due to the labelling degeneracy of the internal (diagonal) momentum. Via the triangle conditions on $\vk_i,\vK$, non-trivial components are specified by
\beq
    q_1\leq q_2,\quad q_3\leq q_4,\quad q_1\leq q_3,\quad |q_1-q_2|\leq Q\leq q_1+q_2, \quad |q_3-q_4|\leq Q\leq q_3+q_4,
\eeq

\subsection{Parity-Odd Estimator}
To estimate the parity-odd trispectrum components, we use a similar scheme, but insert a factor of $\vk_1\cdot\vk_2\times\vk_3$ to pick out only the imaginary component:
\beq
    \widehat{T}_-(q_1,q_2,q_3,q_4,Q) \propto \int_{\vk_i\in q_i,\,\vK\in Q}\left[\prod_{i=1}^4\delta(\vk_i)\right]\delD{\vk_1+\vk_2-\vK}\delD{\vk_3+\vk_4+\vK}(\vk_1\cdot\vk_2\times\vk_3).
\eeq
Note that parity-even contributions are nulled in this estimator due to $\vk\to-\vk$ labelling symmetries. To form a practical estimator, the triple product can be rewritten as a Cartesian sum, leading to:
\beq
    \widehat{T}_-(q_1,q_2,q_3,q_4,Q) &\propto& \epsilon_{ijk}\int_{\vK\in Q}\left[\int d\vx\,\left(\int_{\vk_1\in q_1}k^i_1\delta(\vk_1)e^{-i\vk_1\cdot\vx}\right)\left(\int_{\vk_2\in q_2}k^j_2\delta(\vk_2)e^{-i\vk_2\cdot\vx}\right)e^{i\vK\cdot\vx}\right]\\\nonumber
    &&\,\times\,\left[\int d\vy\,\left(\int_{\vk_3\in q_3}k^k_3\delta(\vk_3)e^{-i\vk_3\cdot\vy}\right)\left(\int_{\vk_4\in q_4}\delta(\vk_4)e^{-i\vk_4\cdot\vy}\right)e^{-i\vK\cdot\vy}\right],
\eeq
This is computed similarly to before, but now requires computing four Fourier transforms per $k$-bin (for $i=1,2,3$ and without $\vk$). The normalization factor is defined as for the even case, without the triple product, though the particular choice is, to an extent, arbitrary.

An alternative (but equivalent) approach to compute the estimator is to expand the triple product in spherical harmonics, via the standard relation
\beq\label{eq: triple-product-Ylm}
    \vk_1\cdot\vk_2\times\vk_3 = -\sqrt{6}i\left(\frac{4\pi}{3}\right)^{3/2}k_1k_2k_3\sum_{m_1m_2m_3}\tj{1}{1}{1}{m_1}{m_2}{m_3}Y_{1m_1}(\hk_1)Y_{1m_2}(\hk_2)Y_{1m_3}(\hk_3).
\eeq
This separates the $\vk_i$ dependence, allowing for the estimator to be written as
\beq
    \widehat{T}_-(q_1,q_2,q_3,q_4,Q) &\propto& -\sqrt{6}i\left(\frac{4\pi}{3}\right)^{3/2}\sum_{m_1m_2m_3}\tj{1}{1}{1}{m_1}{m_2}{m_3}\\\nonumber
    &&\,\times\,\int_{\vK\in Q}\left[\int d\vx\,\left(\int_{\vk_1\in q_1}k_1Y_{1m_1}(\hk_1)\delta(\vk_1)e^{-i\vk_1\cdot\vx}\right)\left(\int_{\vk_2\in q_2}k_2Y_{1m_2}(\hk_2)\delta(\vk_2)e^{-i\vk_2\cdot\vx}\right)e^{i\vK\cdot\vx}\right]\\\nonumber
    &&\quad\,\times\,\left[\int d\vy\,\left(\int_{\vk_3\in q_3}k_3Y_{1m_3}(\hk_3)\delta(\vk_3)e^{-i\vk_3\cdot\vy}\right)\left(\int_{\vk_4\in q_4}\delta(\vk_4)e^{-i\vk_4\cdot\vy}\right)e^{-i\vK\cdot\vy}\right],
\eeq
for $m_i\in\{-1,0,1\}$ and $m_1+m_2+m_3=0$. This again requires four FFTs per $k$-bin, though only a total of seven terms need to be combined together.

The above estimator has various symmetry properties:
\beq
    \widehat{T}_-(q_1,q_2,q_3,q_4,Q) = -\widehat{T}_-(q_2,q_1,q_3,q_4,Q) = \widehat{T}_-(q_3,q_4,q_1,q_2,Q);
\eeq
(noting the negative sign under $q_1\leftrightarrow q_2$ or $q_3\leftrightarrow q_4$ interchange); as before, modes with $\widehat{T}_-(q_1,q_3,q_2,q_4,Q)$ are partially covariant with $\widehat{T}_-(q_1,q_2,q_3,q_4,Q)$, due to the labelling degeneracy of the internal (diagonal) momentum. Finally we note that, if $q_1=q_2$ or $q_3=q_4$, the trispectrum must vanish, based on the above symmetries. This modifies the condition for non-trivial bins to
\beq
    q_1<q_2,\quad q_3<q_4,\quad q_1\leq q_3,\quad |q_1-q_2|\leq Q\leq q_1+q_2, \quad |q_3-q_4|\leq Q\leq q_3+q_4,
\eeq
and one can additionally impose $q_2\leq q_4$ if $q_1=q_3$.

\section{Theoretical Primordial Trispectra}\label{app: tl-theory}

In the below, we consider how to relate the above trispectrum estimators to theoretical models, \textit{i.e.}\ to compute the expectation of $T_{\pm}$ for a given model for $\av{\delta(\vk_1)\cdots\delta(\vk_4)}$, specified by $\tau_\pm$, as in \eqref{eq: Tl-general}. 

\subsection{Parity-Even Trispectrum}

In expectation, the parity-even trispectrum estimator yields:
\beq
    \mathbb{E}[\widehat{T}_+(q_1,q_2,q_3,q_4,Q)] &\propto& \int_{\vk_i\in q_i,\,\vK\in Q}\tau_+(k_1,k_2,k_3,k_4,K,K')\delD{\vk_1+\vk_2-\vK}\delD{\vk_3+\vk_4+\vK}.
\eeq
Assuming $\tau_+$ to be independent of the second diagonal $K'\equiv|\vk_1+\vk_3|$, this can be simplified by rewriting the Dirac deltas as exponentials and using the result $\int_{\hk}e^{-i\vk\cdot\vx} = j_0(kx)$:
\beq
    \mathbb{E}[\widehat{T}_+(q_1,q_2,q_3,q_4,Q)] &\propto& \int_{k_i\in q_i,\,K\in Q}\tau_+(k_1,k_2,k_3,k_4,K)\\\nonumber
    &&\,\times\,\left(4\pi\int x^2dx\,j_0(k_1x)j_0(k_2x)j_0(Kx)\right)\left(4\pi\int y^2dy\,j_0(k_3y)j_0(k_4y)j_0(Ky)\right).
\eeq
Further noting that 
\beq\label{eq: triple-bessel0}
    4\pi \int x^2dx\,j_0(ax)j_0(bx)j_0(cx) = \frac{\pi^2}{abc}\Delta(a,b,c),
\eeq
where $\Delta=1$ if $\{a,b,c\}$ obey triangle conditions and zero else, this can be written
\beq
    \mathbb{E}[\widehat{T}_+(q_1,q_2,q_3,q_4,Q)] &\propto& \int_{k_i\in q_i,\,K\in Q}\frac{\pi^4}{(k_1k_2k_3k_4)K^2}\tau_+(k_1,k_2,k_3,k_4,K)\Delta(k_1,k_2,K)\Delta(k_3,k_4,K),
\eeq
where the factors of $\Delta$ can be dropped if all modes in the bin obey triangle conditions. The normalization is identical, except without the factor of $\tau_+$.

In the thin-bin limit, we can drop the $k_i$ and $K$ integrals, giving a normalization of $q_1q_2q_3q_4(\delta k)^5/(32\pi^6)$, and thus the full trispectrum
\beq\label{eq: even-trispectrum-limit}
    \mathbb{E}[\widehat{T}_+(q_1,q_2,q_3,q_4,Q)] &\approx& \tau_+(q_1,q_2,q_3,q_4,Q).
\eeq
In the limit of a $K'$-independent $\tau_+$, the parity-even trispectrum is thus an unbiased estimator of $\tau_+$ in the thin-bin limit.

\subsection{Parity-Odd Trispectrum}\label{app:TheoryParityOddTris}
A similar procedure can be performed for the parity-odd trispectrum estimator, starting from the general form
\beq\label{eq: tau-phi-def}
    \mathbb{E}[\widehat{T}_-(q_1,q_2,q_3,q_4,Q)] &\equiv&  i\int_{\vk_i\in q_i,\,\vK\in Q}(\vk_1\cdot\vk_2\times\vk_3)^2\tau_-(k_1,k_2,k_3,k_4,K,K')\\\nonumber
    &&\,\times\,\delD{\vk_1+\vk_2-\vK}\delD{\vk_3+\vk_4+\vK}.
\eeq
To simplify this, we first rewrite the squared triple product in spherical harmonics, starting from \eqref{eq: triple-product-Ylm}:
\beq
    (\hk_1\cdot\hk_2\times \hk_3)^2 
    &=& -6(4\pi)^{3/2}\sum_{L_iM_i}Y_{L_1M_1}(\hk_1)Y_{L_2M_2}(\hk_2)Y_{L_3M_3}(\hk_3)\tj{L_1}{L_2}{L_3}{M_1}{M_2}{M_3}\\\nonumber
    &&\times\,\tjo{1}{1}{L_1}\tjo{1}{1}{L_2}\tjo{1}{1}{L_3}\sqrt{(2L_1+1)(2L_2+1)(2L_3+1)}\begin{Bmatrix} L_1 & L_2 & L_3\\ 1 & 1 & 1\\ 1 & 1 & 1\end{Bmatrix},
\eeq
where $L_i\in\{0,2\}$ and we have used the product identity of spherical harmonics as well as the definition of the Wigner $9j$ symbol. Next, we rewrite the Dirac deltas as exponentials and perform the $\hk_i$ integrals via $\int_{\hk}e^{\pm i\vk\cdot\vx}Y_{LM}(\hk) = i^{\pm L} j_L(kx)Y_{LM}(\hx)$. Assuming the trispectrum to be $K'$-independent, as before, this yields
\beq
    \mathbb{E}[\widehat{T}_-(q_1,q_2,q_3,q_4,Q)] &\propto& -6i(4\pi)^{1/2}\int_{k_i\in q_i,\,K\in Q}k_1^2k_2^2k_3^2\,\tau_-(k_1,k_2,k_3,k_4,K)\\\nonumber
    &&\,\times\,\sum_{L_iM_i}\left(4\pi\int x^2dx\,j_{L_1}(k_1x)j_{L_2}(k_2x)j_{L_3}(Kx)\right)i^{-L_1-L_2-L_3}\tj{L_1}{L_2}{L_3}{M_1}{M_2}{M_3}\\\nonumber
    &&\,\times\,\left(\int d\hx\,Y_{L_1M_1}(\hx)Y_{L_2M_2}(\hx)Y_{L_3M_3}(\hx)\right)\left(4\pi \int y^2dy\,j_{L_3}(k_3y)j_0(k_4y)j_{L_3}(Ky)\right)\\\nonumber
    &&\times\,\tjo{1}{1}{L_1}\tjo{1}{1}{L_2}\tjo{1}{1}{L_3}\sqrt{(2L_1+1)(2L_2+1)(2L_3+1)}\begin{Bmatrix} L_1 & L_2 & L_3\\ 1 & 1 & 1\\ 1 & 1 & 1\end{Bmatrix},
\eeq
where we have integrated over $\hy$ then $\hat{\vK}$. All $k_i$ and $K$ integrals are performed with respect to the Lebesgue measure $k^2dk/(2\pi^2)$. Finally, we note that the $\hx$ integral is a Gaunt factor, and use $3j$ completeness to yield the result:
\beq
    \mathbb{E}[\widehat{T}_-(q_1,q_2,q_3,q_4,Q)] &\propto& -6i\int_{k_i\in q_i,\,K\in Q}k_1^2k_2^2k_3^2\,\tau_-(k_1,k_2,k_3,k_4,K)\sum_{L_1L_2L_3}(2L_1+1)(2L_2+1)(2L_3+1)\\\nonumber
    &&\,\times\,i^{-L_1-L_2-L_3}\tjo{L_1}{L_2}{L_3}\tjo{1}{1}{L_1}\tjo{1}{1}{L_2}\tjo{1}{1}{L_3}\begin{Bmatrix} L_1 & L_2 & L_3\\ 1 & 1 & 1\\ 1 & 1 & 1\end{Bmatrix}\nonumber\\
    &&\,\times\,\left(4\pi\int x^2dx\,j_{L_1}(k_1x)j_{L_2}(k_2x)j_{L_3}(Kx)\right)\left(4\pi \int y^2dy\,j_{L_3}(k_3y)j_0(k_4y)j_{L_3}(Ky)\right).
\eeq
This is just a collection of coupled one-dimensional integrals, and the full spectrum can be evaluated with two-dimensional quadrature. We note that the $x$ and $y$ integrals are analytic, taking the form \citep{1991JPhA...24.1435M}:
\beq\label{eq: triple-bessel}
    4\pi\int x^2dx\,j_{\ell_1}(ax)j_{\ell_2}(bx)j_{\ell_3}(cx) &=& \frac{\pi^2}{abc}\Delta(a,b,c)\tjo{\ell_1}{\ell_2}{\ell_3}^{-1}i^{\ell_1+\ell_2-\ell_3}\sqrt{2\ell_3+1}\left(\frac{a}{c}\right)^{\ell_3}\\\nonumber
    &&\,\times\,\sum_{L=0}^{\ell_3}\begin{pmatrix}2\ell_3\\ 2L\end{pmatrix}^{1/2}\left(\frac{b}{a}\right)^L\sum_{\ell'=|\ell_2-L|}^{\ell_2+L}(2\ell'+1)\tjo{\ell_1}{\ell_3-L}{\ell'}\tjo{\ell_2}{L}{\ell'}\\\nonumber
    &&\,\times\,\begin{Bmatrix}\ell_1 & \ell_2 & \ell_3\\ L & \ell_3-L & \ell'\end{Bmatrix} L_{\ell'}\left(\frac{a^2+b^2-c^2}{2ab}\right),
\eeq
where $L_\ell$ is a Legendre polynomial, and $\Delta$ ensures triangle conditions, as before. This involves both $3j$ and $6j$ Wigner symbols, and here requires only $\ell_i\in\{0,2\}$. The $y$ integrals always involve $\ell_3=0$, yielding
\beq
    4\pi\int y^2dy\,j_{\ell}(Ky)j_{\ell}(k_3y)j_{0}(k_4y) &=& \frac{\pi^2}{k_3k_4K}\Delta(k_3,k_4,K)L_\ell\left(\frac{K^2+k_3^2-k_4^2}{2Kk_3}\right),
\eeq
where $\ell\in\{0,2\}$ is required. We note that this makes the separability in $\{k_3,k_4,K\}$ less trivial (though still sum-separable, since the Legendre functions are polynomial).

In the thin-bin limit, we can write
\beq
    \mathbb{E}[\widehat{T}_-(q_1,q_2,q_3,q_4,Q)] &\approx& -\frac{6}{\pi^2}iq_1^3q_2^3q_3^2Q\,\tau_-(q_1,q_2,q_3,q_4,Q)\sum_{L_1L_2L_3}i^{-L_1-L_2-L_3}(2L_1+1)(2L_2+1)(2L_3+1)\nonumber\\
    &&\,\times\,\tjo{L_1}{L_2}{L_3}\tjo{1}{1}{L_1}\tjo{1}{1}{L_2}\tjo{1}{1}{L_3}\begin{Bmatrix} L_1 & L_2 & L_3\\ 1 & 1 & 1\\ 1 & 1 & 1\end{Bmatrix}\nonumber\\
    &&\,\times\,\left(4\pi\int x^2dx\,j_{L_1}(q_1x)j_{L_2}(q_2x)j_{L_3}(Qx)\right)L_{L_3}\left(\frac{Q^2+q_3^2-q_4^2}{2Qq_3}\right),
\eeq
assuming triangle conditions to be satisfied. This is much more complex than the parity-even equivalent \eqref{eq: even-trispectrum-limit}, due to the $(\vk_1\cdot\vk_2\times\vk_3)^2$ factor. If this was additionally included in the normalization, the limit would be simply $i\tau_-(q_1,q_2,q_3,q_4,Q)$; however, this is likely to be unstable for close-to-coplanar tetrahedral configurations, whence $(\vk_1\cdot\vk_2\times\vk_3)^2$ vanishes.

\bibliographystyle{apsrev4-2}
\bibliography{refs}

\end{document}